\documentclass[aps,pre,twocolumn,amsmath,amssymb,amsfonts]{revtex4}
\usepackage{epsfig}
\usepackage{graphicx}
\usepackage{dcolumn}
\usepackage{bm}
\usepackage{enumitem}
\usepackage{xcolor}

\usepackage{bm}
\usepackage{xcolor}

\usepackage{color}
\usepackage{ulem}
\usepackage{bm}
\usepackage{soul, xcolor}
\usepackage{xcolor,cancel}
\usepackage[titletoc,title]{appendix}

\definecolor{amber}{rgb}{1.0, 0.75, 0.0}
\definecolor{bittersweet}{rgb}{1.0, 0.44, 0.37}
\definecolor{cadmiumorange}{rgb}{0.93, 0.53, 0.18}

\begin{document}
\setstcolor{red}
\title{Linear and nonlinear hydromagnetic stability in laminar and turbulent flows}
\author{Itzhak Fouxon$^{1,2}$}\email{itzhak8@gmail.com}
\author{Joshua Feinberg$^{1}$}\email{jfeinberg@univ.haifa.ac.il}
\author{Michael Mond$^2$}\email{mondmichael@gmail.com}
\affiliation{$^1$ Department of Mathematics and Haifa Research Center for Theoretical Physics and Astrophysics, University of Haifa,
Haifa 31905, Israel}
\affiliation{$^2$ Department of Mechanical Engineering, Ben-Gurion University of the Negev, Beer Sheva 84105, Israel}

\begin{abstract}

We consider the evolution of arbitrarily large perturbations of a prescribed pure hydrodynamical flow of an electrically conducting fluid.
We study whether the
flow perturbations as well as the generated magnetic fields decay or grow with time and constitute a dynamo process.
For that purpose we derive a generalised Reynolds-Orr equation for the sum of the kinetic energy of the
hydrodynamic perturbation and the magnetic energy. The flow is confined in a finite volume so the normal
component of the velocity at the boundary is zero. The tangential component is left arbitrary in contrast with previous works. For the magnetic field we mostly employ the classical boundary conditions where the field
extends in the whole space.
We establish critical values of hydrodynamic and magnetic Reynolds numbers below which arbitrarily large initial perturbations of the hydrodynamic flow decay. This involves generalisation of the Rayleigh-Faber-Krahn
inequality for smallest eigenvalue of an elliptic operator. For high Reynolds number turbulence we provide an estimate of critical
magnetic Reynolds number below which arbitrarily large fluctuations of the
magnetic field decay.


\end{abstract}

\maketitle

\section{Introduction}
Flows of electrically conducting fluids are of major interest in fluid mechanics with a vast array of applications including plasmas, liquid metals and salt water. In many cases no source of significant magnetic fields is
present so that the field can be maintained only due to the energy transfer between the flow and the field. The main mechanism of the transfer in incompressible case consists of stretching of the magnetic field lines \citep{ll8,david}, starting at
some point in the past with a seed, possibly small, magnetic field into which the flow started to pump energy.  It seems inevitable, by analogy with ordinary fluid mechanical stability \citep{ll8,drazin} that, for dissipation
coefficients of viscosity and magnetic diffusivity larger than some critical values, the laminar hydrodynamic flow is stable. This fundamental property, not limited to the linear stability with respect to small disturbances,
is proved in fluid mechanics by the study of the so-called Reynolds-Orr equation equation \citep{ll8,drazin}. However seemingly no similar equation has been written in magnetohydrodynamics (MHD), see e. g. \cite{david}. Here we fill
in this gap and provide different extensions for the cases of small and large disturbances. We demonstrate that the equation is as useful as its hydrodynamic counterpart and also provides a tool for calculating the logarithmic
rate of change of the magnetic field.

\section{Overview of the main results}
We start with highlighting the main results to be unfolded below.

The problem is formulated and discussed in Section \ref{adma} while an extended Reynolds-Orr equation is derived in Section \ref{admb} as a measure of the magnitude of the perturbations. It should be stressed that it is not assumed that the perturbations are small. Equation (\ref{xs}) is the main result of Section \ref{adm} and one of the main results of the current paper, and serves as a starting point and basis for the rest of the calculations.

Global stability criteria are provided in Section \ref{stability}. Thus, stability criteria in terms of the stress tensor is presented in Section \ref{stabilitys} in Eqs. (\ref{crt}) while in Section \ref{glo} alternative criteria are derived in terms of the kinetic and magnetic Reynolds numbers, and are given in Eq. (\ref{criteria}).

Section \ref{variational} provides proofs for the criteria given in Eqs. (\ref{crt}) and (\ref{criteria}). It contain however much more than that, as en-route to proving the necessary inequalities several novel results of general interest are obtained.  In particular, a generalised Rayleigh-Faber-Krahn inequality is derived, based on the extension of the Rayleigh conjecture to solenoidal vectors, as given by Eq. (\ref{lambda1}). The inequality is derived by solving the vector eigenvalue problem (\ref{op}).  Another new inequality is given in Eq. (\ref{re1}). The relevance to free decaying hydrodynamic flows and magnetic fields is demonstrated and discussed. In particular, it is demonstrated in Eq. (\ref{zerop}) that the slowest decaying hydrodynamical mode is of zero pressure.

Section \ref{obc} addresses the impact of replacing the true physical boundary conditions, that are employed in the current work, with those frequently used in computational astrophysics, namely either the normal component of the magnetic field or its tangential component are zero on the computational domain boundaries. The answer ia given in Eq. (\ref{soot}) which demonstrates that the numerical boundary conditions slightly underestimate the value of the electric diffusivity needed for stabilising the system.

Section \ref{tr} is devoted to the derivation of the lower threshold on the magnetic Reynolds number for the turbulent flow where the kinetic Reynolds number is high. The threshold, below which the magnetic field of arbitrary initial magnitude decays, is obtained by the requirement of self-consistency of assumption of the magnetic field growth. The next Section studies the growth of magnetic field in turbulence in the kinematic regime where the magnetic field is small and can be neglected in the Navier-Stokes equations. We introduce refinements of Childress and Backus bounds. These are of interest already in the case of laminar flow. In the case of turbulence, using the assumption that the magnetic field spectrum is separable, we provide a detailed estimate on the growth exponent. Conclusions Section summarizes the main results of the work and Appendix provides some technical details of the calculations.

\section{Energy equation for non-linear dynamo problem} \label{adm}
Magnetic disturbances of large and small magnitudes must be distinguished. For small initial disturbances the Lorentz force that describes the impact of the magnetic field on the flow is negligible.
Therefore the hydrodynamic flow initially obeys closed evolution which is independent of the magnetic field. The evolution of the magnetic field is passive. It is described by the induction equation which is linear in the field and resembles the linearised Navier-Stokes equations from the usual stability problem of the hydrodynamic flow. The question of whether the field grows, eventually creating a large Lorentz force that can no longer be neglected, is studied by the
kinematic dynamo problem. However, even if it is found that the magnetic field decays, this does not guarantee that the hydrodynamic flow, defined here as the flow with zero magnetic field, is realised.
The linear stability of small initial perturbations guarantees only that the magnetic field decays at large times, allowing for growth at intermediate times where the field can become so large as to produce significant impact on the flow and lead to a non-linear instability \citep{roberts}. Thus linear stability analysis is incomplete even in the case of small initial perturbations of the magnetic field. If the initial disturbances are already large, non-linear instability can occur directly.
We start from the study of this non-linear or global stability.

\subsection{Magnetohydrodynamic equations and admissible fields}\label{adma}

We consider the magnetohydrodynamic (MHD) equations of incompressible flow $\bm v$ coupled to the (rescaled) solenoidal magnetic field $\bm B$, see e.g. \cite{ll8},
\begin{eqnarray}&&\!\!\!\!\!\!\!\!\!\!\!\!\!\!
\partial_t\bm v+(\bm v\cdot \nabla) \bm v=-\nabla p+\nu \nabla^2\bm v+(\bm B\cdot \nabla)\bm B+\bm f, \label{magnc} \\&&\!\!\!\!\!\!\!\!\!\!\!\!\!\!
\partial_t\bm B+(\bm v\cdot \nabla) \bm B=(\bm B\cdot \nabla) \bm v+\eta\nabla^2 \bm B,\ \ \nabla\cdot\bm v=0. \nonumber
\end{eqnarray}
The flow is driven by the body forces $\bm f$ and/or boundary condition with a prescribed flow at the boundary of the region $S$. The fluid density is set to one so that $p$ is the sum of the hydrodynamic and magnetic pressure terms, $\nu$ is the kinematic viscosity and $\eta$ is the magnetic diffusivity. We assume that the fluid occupies a fixed finite domain with volume $V$ so that normal component of the fluid velocity $v_n\equiv \bm v\cdot \bm {\hat n}$ is zero on $S$, where $\bm {\hat n}$ is the outward-pointing normal. In the domain's exterior $V'$ there is no electric current and the magnetic field is potential, $\nabla\times\bm B=0$. The remaining boundary conditions are continuity of the solenoidal field $\bm B$ across $S$ and the requirement that $r^3B$ is bounded in the whole space. These boundary conditions for Eqs.~(\ref{magnc}) lead to a unique solution of Maxwell's equations \cite{backus}.

We consider a constraint on the admissible initial conditions for Eqs.~(\ref{magnc}) otherwise understood as the characterisation of the functional spaces to which $\bm v$ and $\bm B$ belong. The only general constraint on $\bm v$ is that it is solenoidal and equals the prescribed boundary condition on $S$. The constraint on admissible $\bm B$ that could serve as a possible initial condition for Eqs.~(\ref{magnc}) is more delicate. The field continuity on $S$ gives the requirement that the limiting value of the magnetic field $\bm B$ on $S$ approached from inside $V$ equals $\nabla\phi$ where $\phi$ is the harmonic potential that provides the magnetic field outside $V$ via $\bm B=\nabla \phi$. This condition can be given an explicit form by observing that $\phi$, which we assume to vanish at infinity, is defined uniquely by its normal derivative $\bm {\hat n}\cdot\nabla\phi(S)$ on the boundary providing the Neumann boundary condition. Therefore, since the problem is linear, then $\phi(\bm x)$ with $\bm x$ in $V'$ must be given by a certain linear integral transformation of the boundary condition:
\begin{eqnarray}&&\!\!\!\!\!\!\!\!\!\!\!\!\!\!
\phi(\bm x)=-\int_S G(\bm x, \bm x') B_n(\bm x')dS,\ \ B_n\equiv 
 \bm {\hat n}\cdot\nabla\phi(S), \label{ref}
\end{eqnarray}
where $G(\bm x, \bm x')$ is the transformation's kernel and $B_n$ is the normal component of the magnetic field on the surface. The kernel is the fundamental solution (Green's function) of the Laplace equation on $V'$ with the Neumann boundary condition. Its explicit form is known only for simplest domains, such as the case where $V$ is a ball, see e.g. \cite{green,free}. We introduced the minus sign in the definition both since $B_n$ is the normal derivative with the normal inside $V'$ (and not outside as is more usual in the definition) and for future convenience. The domain of variation of $\bm x'$ in $G(\bm x, \bm x')$ is $S$ though continuation outside $S$ can be considered with no difficulty.

It is useful to observe that the restriction of $G(\bm x, \bm x')$ to $\bm x$ on $S$ is a symmetric function of $\bm x$ and $\bm x'$. This can be proved by considering two harmonic functions $G(\bm x, \bm x_1)$ and $G(\bm x, \bm x_2)$ on $V'$ for arbitrary $\bm x_1$ and $\bm x_2$ on $S$. By definition the normal derivatives of these solutions on $S$ are surface $\delta-$functions concentrated at $\bm x_1$ and $\bm x_2$, respectively. Then using Green's second identity is readily seen to give $G(\bm x_2, \bm x_1)=G(\bm x_1, \bm x_2)$ proving the symmetry.

The Green's function $G(\bm x, \bm x')$ can be continued with no difficulty to $\bm x$ in $V$, see e.g. the explicit formulas for the ball in \cite{green,free}. Therefore the representation of the potential given by Eq.~(\ref{ref}) also provides an explicit continuation of $\phi$ inside $V$. The continuation is strongly non-unique and yet it is useful for calculations, see \cite{backus} and below. The continuation's details are irrelevant in the final formulas.

The Green's function provides insight into the magnetic field energy $E_e$ contained outside the flow domain. Thus, expressing $B^2$ in $V'$ as $B^2=\nabla\cdot(\phi\nabla \phi)$ we obtain:
\begin{eqnarray}&&\!\!\!\!\!\!\!\!\!\!\!\!\!\!
E_e\!\equiv\! \int_{V'} \!\!\frac{B^2}{2}d\bm x\!=\!\frac{1}{2}\!\int_{V'} \!\!\nabla\!\cdot\!(\phi\nabla \phi)d\bm x\!=\!-\frac{1}{2}\!\int_S \phi B_n dS
\nonumber\\&&\!\!\!\!\!\!\!\!\!\!\!\!\!\!
\!=\!\frac{1}{2}\!\int B_n(\bm x) G(\bm x, \bm x')B_n(\bm x')dS dS'. \label{defe}
\end{eqnarray}
Thus for arguments on the surface the symmetric function $G(\bm x, \bm x')$ can be interpreted as the "potential" energy of interaction of different surface parts. It is seen from the last equation that $G(\bm x, \bm x')$, restricted to arguments on the surface, is a positive definite symmetric kernel. In other words it is a real Hermitian operator with positive eigenvalues which might be of use in some applications.

Differentiating $\phi$ as given by Eq.~(\ref{ref}), and employing the continuity of $\bm B$ across $S$ produce an expression for the tangential components of $\bm B$ on $S$ via $B_n(S)$
\begin{eqnarray}&&\!\!\!\!\!\!\!\!\!\!\!\!\!\!
B_i(S)=-\left(\int_S \frac{\partial G(\bm x, \bm x')}{\partial x_i} B_n(\bm x')dS\right)_{\bm x\to S},
\label{constraint}
\end{eqnarray}
where the limit is taken from $V'$. This equation gives a constraint on the limiting values of $\bm B(S)$ approached either from $V$ or $V'$. It provides the tangential components of $\bm B(S)$ via the normal component $B_n(S)$.
Conversely, given a vector field $\bm B$ defined on $V$ and obeying Eq.~(\ref{constraint}) we can construct admissible solution for $\bm B$ in the whole space by continuing $\bm B$ to $V'$ via $\bm B=\nabla\phi$ and Eq.~(\ref{ref}). It is readily seen by multipole-type expansion of $G(\bm x, \bm x')$ in Eq.~(\ref{ref}) that at large $\bm x$ we have in the leading order
\begin{eqnarray}&&\!\!\!\!\!\!\!\!\!\!\!\!\!\!
\phi(\bm x)\sim -\int_S x'_i \frac{\partial G(\bm x, \bm x') }{\partial x'_i}|_{\bm x'=0} B_n(\bm x')dS,
\end{eqnarray}
where we used that $\int_S B_n(\bm x')dS=0$ by $\nabla\cdot\bm B=0$. The behavior of Green's function at large distances implies that the resulting $\bm B$ obeys the condition that $r^3B$ is bounded in the whole space, thus finishing the construction of globally admissible $\bm B$.

The above observations allow to confine the theoretical and numerical studies to $V$, by using the no-slip boundary condition for $\bm v$, as well as the boundary condition (\ref{constraint}) for $\bm B$, on $S$. For instance in the case where $V$ is a ball, which provides a reasonable description of the liquid core of the earth and other cores, we can use Eq.~(\ref{constraint}), with explicit formula for $G(\bm x, \bm x')$ provided in \cite{green,free}, as the boundary condition on $\bm B$ and confine the study to $V$. For simulations of generic domains numerical generation of $G(\bm x, \bm x')$ would be needed.

The setting provided by Eqs.~(\ref{magnc}), and the boundary conditions that are laid out below it, describes situations similar to conducting liquid in the core of the earth and a number of astrophysical situations. However, numerical simulations are often confined to the volume of the fluid $V$ and use a different set of boundary conditions on $\bm B$. Periodic boundary conditions are used often and they in fact simplify the calculations below. Other boundary conditions in use are zero normal or tangential components of the magnetic field on $S$. We will perform the calculations by using the framework described after Eqs.~(\ref{magnc}) which is probably more reasonable
physically. Then we will comment how the calculations change for the periodic or other boundary conditions in the simulations.

\subsection{Perturbation magnitude}\label{admb}

The MHD equations allow for a purely hydrodynamic flow solution, $\bm B\equiv 0$, where the flow $\bm v_0$ and the pressure $p_0$ solve the incompressible Navier-Stokes equations
\begin{eqnarray}&&\!\!\!\!\!\!\!\!\!\!\!\!\!\!
\partial_t\bm v_0+(\bm v_0\cdot \nabla) \bm v_0=-\nabla p_0+\nu \nabla^2\bm v_0+\bm f,\ \ \nabla\cdot\bm v_0=0.\label{nc}
\end{eqnarray}
We are searching for a condition that guarantees that any initial disturbance $\bm v(t=0)$ and $\bm B(t=0)$, whose evolution is governed by Eqs.~(\ref{magnc}), decays to $\bm v_0$. The linear stability analysis does not apply since the disturbances are not assumed to be small. We observe that $\bm u=\bm v-\bm v_0$ obeys
\begin{eqnarray}&&
\partial_t\bm u+(\bm v\cdot \nabla) \bm u+(\bm u\cdot \nabla) \bm v_0=\nabla(p_0-p)+\nu \nabla^2\bm u
\nonumber\\&& +(\bm B\cdot \nabla)\bm B; \ \ \bm u(S)=0,\ \ \nabla\cdot\bm u=0.
\label{uequation}
\end{eqnarray}
Scalar multiplication with $\bm u$ and integration gives for the ``kinetic energy" (this is not a true kinetic energy since $\bm u$ is not a flow however we will often speak of it below as kinetic energy) $E_{kin}\equiv \int_V u^2d\bm x/2$ that
\begin{eqnarray}&&\!\!\!\!\!\!\!\!\!\!\!\!\!\!
\dot{E}_{kin}\!=\!-\int_V\!\! \left(\bm u s\bm u\! +\!\nu w^2\!+\!\bm B(\bm B\!\cdot\! \nabla)\bm u\right) d\bm x,\nonumber\\&&\!\!\!\!\!\!\!\!\!\!\!\!\!\!
 s_{ik}\equiv \frac{\nabla_iv_{0k}+\nabla_kv_{0i}}{2},\ \ \bm w\equiv \nabla\times \bm u, \label{soi1}
\end{eqnarray}
where $s$ is the rate-of-strain tensor and we used $\bm u(S)=0$. We introduced the vorticity $\bm w$ which we sometimes write below explicitly as $\nabla\times \bm u$. We rewrote the dissipation term in the standard way by using the identity $(\nabla\bm f)^2-\nabla \cdot((\bm f\cdot\nabla)\bm f)=(\nabla\times \bm f)^2$
that holds for any solenoidal vector field $\bm f$. Integration of this identity for $\bm f=\bm u$ gives $\int_V \left(\nabla \bm u\right)^2 d\bm x=\int_V \left(\nabla\times \bm u\right)^2 d\bm x$ due to the condition $\bm u(S)=0$.

In contrast with the kinetic energy of the fluid, the magnetic energy density is present in the whole space $V+V'$. Thus the total energy $E_m$ is the sum $E_m=E_i+E_e$ of the interior energy $E_i\equiv \int_{V} B^2d\bm x/2$ and the exterior energy $E_e\equiv \int_{V'} B^2d\bm x/2$. Calculation of the rate of change of $E_m$ was done in \cite{gellman,backus}. Here we reproduce the calculation both for later reference to the equations and because there are some differences. Thus the calculations of \cite{gellman} have been criticized in \cite{backus} for the unjustified use of Poynting's theorem. Here we perform the calculation along the lines of \cite{backus} however without assuming $\bm v(S)=0$, we assume only that the normal component is zero, namely $v_n(S)=0$.

{\bf Interior energy}---It is useful to write the diffusivity term as $\nabla^2\bm B=-\nabla\times\bm j$ where $\bm j\equiv \nabla\times \bm B$ is the electric current density. We find by using the identity $\bm B\cdot (\nabla\times\bm j) =\nabla\cdot (\bm j\times \bm B)+j^2$ that
the rate of change of the magnetic energy contained inside the volume of the fluid is
\begin{eqnarray}&&\!\!\!\!\!\!\!\!\!\!\!\!\!\!
\frac{dE_i}{dt}=\frac{d}{dt} \int_V \frac{B^2}{2}d\bm x=\int_V \left(\bm B(\bm B\cdot \nabla)\bm v\! -\!\ \eta j^2\right)d\bm x
\nonumber\\&&\!\!\!\!\!\!\!\!\!\!\!\!\!\!
+\eta \int_S \bm {\hat n} \cdot (\bm B\times \bm j) dS, \label{magnetoc}
\end{eqnarray}
where we used $v_n=0$. The volume term is the energy change due to the flow's stretching of the field lines minus Joule heating. The significance of the surface term can be understood by writing it as a sum of the integral of Poynting's vector and the term due to the flow, see detailed discussion in \cite{gellman}.

{\bf Exterior energy}---We consider the time derivative of the magnetic energy $E_e$ contained outside $V$ which is defined in Eq.~(\ref{defe}). There the magnetic field can be described by a harmonic potential $\bm B=\nabla \phi$. This component of the magnetic field is fully dictated by the field in $V$ so that the rate of exterior  energy change may be expressed in terms of an energy flux through $S$
\begin{eqnarray}&&\!\!\!\!\!\!\!\!\!\!\!\!\!\!
\frac{dE_e}{dt}=\int_{V'} \nabla\phi\cdot \partial_t\bm B dV=\int_{V'} \nabla\cdot \left(\phi\partial_t\bm B\right) dV
\nonumber\\&&\!\!\!\!\!\!\!\!\!\!\!\!\!\!
=-\int_S \phi \partial_t\bm B\cdot \bm {\hat n} dS.
\end{eqnarray}
Here continuity of $\partial_t\bm B\cdot \bm {\hat n}$ allows not to specify which side of the surface is used for defining the limiting value. We therefore find that by continuing $\phi$ smoothly to $V$, as explained after Eq.~(\ref{ref}), the integral can be expressed in terms of the magnetic fields inside $V$. Starting from the last term in the equation above,
\begin{eqnarray}&&\!\!\!\!\!\!\!\!\!\!\!\!\!\!
\frac{dE_e}{dt}=-\int_{V} \nabla\cdot \left(\phi\partial_t\bm B\right) dV=-\int_V \nabla\phi \cdot \partial_t\bm BdV.
\end{eqnarray}
Writing now the induction equation in the form $\partial_t\bm B=\nabla\times (\bm v\times \bm B-\eta\bm j)$ and using $\nabla\phi \cdot \nabla\times (\bm v\times \bm B-\eta\bm j)=\nabla\cdot\left(\nabla\phi\times (\eta\bm j-\bm v\times \bm B)\right)$ we find that
\begin{eqnarray}&&\!\!\!\!\!\!\!\!\!\!\!\!\!\!
\frac{dE_e}{dt}=\int_V \nabla\cdot\left(\nabla\phi\times (\bm v\times \bm B-\eta\bm j)\right) dV
\nonumber\\&&\!\!\!\!\!\!\!\!\!\!\!\!\!\!
=\int_S \left(\bm B\times (\bm v\times \bm B-\eta\bm j)\right)\cdot \bm {\hat n} dS,
\end{eqnarray}
where we used the continuity of $\bm B$ at the surface for writing $\nabla\phi(S)=\bm B$. It should be emphasised again that $\nabla\phi$ in the volume integral is not equal to the magnetic field inside $V$ and coincides with the magnetic field only on the surface $S$. The fields on $S$ are evaluated by approaching the surface from $V$ which reduces $\dot E_e$ to quantities pertaining only to the interior. Finally using $v_n=0$
\begin{eqnarray}&&\!\!\!\!\!\!\!\!\!\!\!\!\!\!
\frac{dE_e}{dt}=-\int_S (\bm v\cdot \bm B)(\bm {\hat n} \cdot  \bm B) dS -\eta\int_S \bm {\hat n} \cdot \left(\bm B\times \bm j\right) dS,\label{magnetoc1}
\end{eqnarray}
where the only difference from \cite{backus} is that we keep the first term on the RHS.

{\bf Full energy}---Finally we find on summing Eqs.~(\ref{magnetoc}) and (\ref{magnetoc1}) that
\begin{eqnarray}&&\!\!\!\!\!\!\!\!\!\!\!\!\!\!
\frac{dE_m}{dt}=\frac{d}{dt} \int \frac{B^2}{2}d\bm x=\int_V \left(\bm B(\bm B\cdot \nabla)\bm v\! -\!\eta j^2\right)d\bm x
\nonumber\\&&\!\!\!\!\!\!\!\!\!\!\!\!\!\!
-\int_S (\bm v\cdot \bm B)(\bm {\hat n} \cdot  \bm B) dS. \label{magneto}
\end{eqnarray}
This is the form found in \cite{gellman}. It could also be found from Faraday's law as pointed out by anonymous referee. The term $\bm B(\bm B\cdot \nabla)\bm v$ describes the change in the magnetic energy due to the stretching of the magnetic field lines in the volume of the fluid. The stretching occurs also on the boundary $S$ where the flow gradients are non-zero. This changes the magnetic energy in the fluid exterior by changing $B_n$, see Eq.~(\ref{defe}).

The surface term in the equation above can be written in a form that looks as stretching of magnetic field lines in $V'$ by continuation of velocity field to $V'$ which will be designated by ${\bm v'}$. The continuation is not unique and yet useful similarly to continuation of $\phi$ above (there is of course no physical flow in $V'$). We observe that for any solenoidal $ {\bm v'}$ that vanishes at infinity and equals $\bm v$ on $S$ we have
\begin{eqnarray}&&\!\!\!\!\!\!\!\!\!\!\!\!\!\!
\int_{V'}\!\! B_iB_k\nabla_i v_k' d\bm x=\int_{V'}\!\!\left( \nabla_i(B_iB_k v_k')- {\bm v'}(\bm B\cdot \nabla)\bm B  \right)d\bm x
\nonumber\\&& \!\!\!\!\!\!\!\!\!\!\!\!\!\!=-\int_S (\bm v\cdot \bm B)(\bm {\hat n} \cdot  \bm B) dS
-\frac{1}{2}\int_{V'}\!\!  \nabla\cdot\left(  {\bm v'} B^2\right) d\bm x\nonumber\\&& \!\!\!\!\!\!\!\!\!\!\!\!\!\!
=-\int_S (\bm v\cdot \bm B)(\bm {\hat n} \cdot  \bm B) dS.\label{st}
\end{eqnarray}
We used that $\bm B$ is solenoidal; $(\bm B\cdot \nabla)\bm B=\nabla B^2/2$ in $V'$ due to $\bm B=\nabla\phi$; and $ {\bm v'}(S)=\bm v_0(S)=\bm v(S)$ has zero normal component. We conclude that we can rewrite Eq.(\ref{magneto}) as
\begin{eqnarray}&&\!\!\!\!\!\!\!\!\!\!\!\!\!\!
\frac{dE_m}{dt}\!=\!\int_V \!\!\left(\bm B(\bm B\!\cdot\! \nabla)\bm v\! -\!\eta j^2\right)d\bm x\!+\!\int_{V'}\!\! \bm B(\bm B\!\cdot \!\nabla) {\bm v'}d\bm x. \label{un}
\end{eqnarray}
This way of writing describes uniformly $V$ and $V'$ where the integral contributions of both domains have identical form ($j=0$ in $V'$). The form of $ {\bm v'}$ is strongly non-unique because there are many solutions to $\nabla\cdot  {\bm v'}=0$ with boundary conditions $ {\bm v'}(S)=\bm v(S)$ and $ {\bm v'}(\infty)=0$.
Finally we observe that Eq.~(\ref{magneto}) can be rewritten with surface integrals by using divergence theorem
\begin{eqnarray}&&\!\!\!\!\!\!\!\!\!\!\!\!\!\!
\frac{dE_m}{dt}=-\int_V \left(\bm v(\bm B\cdot \nabla)\bm B\! +\!\eta j^2\right)d\bm x. \label{start}
\end{eqnarray}
It should be noted that the above equation may be derived directly from applying Faraday's law to the entire space \footnote{We thank an anonymous reviewer for pointing out this derivation.}.
We find by using the boundary condition $\bm v_0(S)=\bm v(S)$ and Eq.~(\ref{soi1}) that the rate of change of the total energy $E={E}_{kin}+E_m$ (here $E$ is more a distance between two solutions of the MHD equations, $\bm v_0$,
$\bm B=0$ and $\bm v$, $\bm B$. The distance is the sum of the $L_2-$norms of the solutions' difference $\bm v-\bm v_0$ and $\bm B$. We refer to it as energy similarly to ${E}_{kin}$ above though it rather characterizes the magnitude of the perturbation) is
\begin{eqnarray}&&\!\!\!\!\!\!\!\!\!\!\!\!\!\!
\frac{dE}{dt}\!=\!\int_V\!\! \left(\bm B s\bm B-\bm u s\bm u\! -\!\nu \left(\nabla\times \bm u\right)^2\!-\!\eta (\nabla\times\bm B)^2\right) d\bm x
\nonumber\\&& \!\!\!\!\!\!\!\!\!\!\!\!\!\!
-\!\int_S \!\!(\bm v_0\!\cdot\! \bm B)(\bm {\hat n} \!\cdot\!  \bm B) dS
\!=\!-\int_V\!\! \left(\bm v_0(\bm B\!\cdot\! \nabla)\bm B\!-\!\bm v_0(\bm u\!\cdot\! \nabla)\bm u
\right.\nonumber\\&& \!\!\!\!\!\!\!\!\!\!\!\!\!\!\left.
+\nu \left(\nabla\times \bm u\right)^2 \!+\!\eta (\nabla\times\bm B)^2\right) d\bm x,\label{xs}
\end{eqnarray}
where both, equivalent, forms of the derivative can be of use. The last form demonstrates the similarity between the rates of change of the energies associated with $\bm u$ and $\bm B$. The first two terms in the integrand can be interpreted as the local energy exchange with the perturbed flow $\bm v_0$, with opposite signs for $\bm u$ and $\bm B$, while the last two terms describe the local energy dissipation. We find that the rate of change is given by a sum of two rather similar quadratic forms in $\bm u$ and $\bm B$
\begin{eqnarray}&&\!\!\!\!\!\!\!\!
\frac{dE}{dt}\!=\!Q_1(\bm u)\!+\!Q_2(\bm B);\ \  Q_1(\bm \xi)\!\equiv\! T(\bm \xi)\!-\!\nu D(\bm \xi),
\nonumber\\&& \!\!\!\!\!\!\!\!
Q_2\!\equiv\! -T'(\bm \xi)\!-\!\eta D(\bm \xi),\ \ D\!\equiv\! \int_V\!\! \left(\nabla\!\times \!\bm \xi\right)^2d\bm x,
\label{landau}
\end{eqnarray}
where we defined two quadratic forms $Q_i(\bm \xi)$ for any solenoidal vector field $\bm \xi$ and introduced
\begin{eqnarray}&&\!\!\!\!\!\!\!\!
T'(\bm \xi)\!=\!T(\bm \xi)\!+\!\int_S \!\!(\bm v_0\!\cdot\! \bm \xi)(\bm {\hat n}\! \cdot \! \bm \xi) dS\!=\!\int_V\!\!\bm v_0(\bm\xi\!\cdot \!\nabla)\bm \xi d\bm x,\nonumber\\&&\!\!\!\!\!\!\!\!
T\equiv \!-\!\int_V\!\! \bm \xi s \bm \xi d\bm x.
\end{eqnarray}
Equations (\ref{xs})-(\ref{landau}) are one of the main results of this paper. Since Eq.~(\ref{xs}) reduces to the Reynolds-Orr equation of hydrodynamic stability for $\bm B\equiv 0$ \citep{ll6,drazin,serrin}, we call it the extended Reynolds-Orr equation.

The main usage of the hydrodynamic Reynolds-Orr equation is that it implies that for Reynolds number below a certain threshold the flow perturbations of arbitrary magnitude decay. This conclusion on the stability is reached in a universal way which is independent of the details of the problem. Thereby the equation is essentially the only way for dealing with non-linear stability due to the general unsolvability of non-linear equations.  Moreover the equation is also a useful shortcut to the linear stability analysis which is often intractable. The main drawback of the approach is that the threshold is typically an order of magnitude too conservative than the actual one. Yet the insight it provides is unique  \citep{ll6,drazin,serrin}.

The usage of the extended Reynolds-Orr equation is similar. We observe that $T$ and $T'$ are linear in the unperturbed flow $\bm v_0$ and $D$ is a non-negative form. This implies that the extended Reynolds-Orr equation reveals the fundamental stability of purely hydrodynamic solutions of the MHD in the limit of large dissipation coefficients.  Significant part of the paper below is devoted to quantitative assessment of the resulting stability bounds.

{\bf Comparison of the usual and extended Reynolds-Orr equations}---
The forms $T$ and $D$ were introduced in \cite{ll8} (see a problem set) where the equivalent form $D(\bm u)\equiv \int \left(\nabla\!\bm u\right)^2d\bm x$ was used for $D$. The form $T'$ is very similar to $T$, differing from it by the boundary term only which is due to the difference that kinetic energy of the fluid is confined to $V$ and magnetic energy is everywhere.
The $T$ and $T'$ terms in $Q_1$ and $Q_2$, respectively, describe the energy transfer from $\bm v_0$ to $\bm u$ and $\bm B$ and are not sign definite. The $D$ term describes the decay of the kinetic and magnetic energy due to viscosity and diffusivity respectively and is positive unless $\bm \xi$ is constant. Eq.~(\ref{landau}) allows to give a simple criterion of global stability - if both $Q_i(\bm \xi)$ are negative for any admissible $\bm \xi$ then the disturbance decays. This way to deriving stability condition was well-studied in the framework of the Reynolds-Orr equation \cite{ll6,serrin,drazin}. In the next section we provide the generalization to our case of MHD that demonstrates general stability when dissipative coefficients $\nu$ and $\eta$ are large.

\section{Global stability at supercritical dissipative coefficients}\label{stability}

In this section we demonstrate that Eqs.~(\ref{landau}) imply generally that any initial conditions, with $\bm B$ however large, decay to the purely hydrodynamic solution $\bm v_0$ if $\nu$ and $\eta$ exceed certain critical values.

We introduce dimensionless variables by rescaling all flows and $\bm B$ by a typical velocity value $V_0$, coordinates by a typical scale $L$ and time by $L/V_0$. Denoting the dimensionless variables by the same letters with no ambiguity, we can write
\begin{eqnarray}&&\!\!\!\!\!\!\!\!\!\!\!\!\!\!
Q_1(\bm u)
=-\frac{D(\bm u)}{Re} \left(1-\frac{Re\  T(\bm u)}{D(\bm u)}\right),\nonumber\\&&\!\!\!\!\!\!\!\!\!\!\!\!\!\! Q_2(\bm B)
=-\frac{D(\bm B)}{Re_m}\left(1+\frac{Re_m\ T'(\bm B)}{D(\bm B)}\right),
\end{eqnarray}
and Eq.~(\ref{landau}) is unchanged. We introduced hydrodynamic and magnetic Reynolds numbers by $Re\equiv V_0L/\nu$ and $Re_m\equiv V_0L/\eta$ respectively. At this stage of the general discussion $V_0$ and $L$ are not specified yet. They will be assigned a more definite meaning in the next two subsections.
The main observation, see e.g. \cite{ll6}, is that the (Rayleigh-type) quotients in the brackets have finite extrema over the admissible fields so that we can introduce the following finite positive quantities:
\begin{eqnarray}&&\!\!\!\!\!\!\!\!\!\!\!\!\!\!
\frac{1}{{\tilde Re}}=max_{\bm u}\frac{T(\bm u)}{D(\bm u)},\ \ \frac{1}{{\tilde Re}_m}=max_{\bm B}\frac{-T'(\bm B)}{D(\bm B)}, \label{dasd}
\end{eqnarray}
where the maxima are taken over solenoidal $\bm u$ that obey $\bm u(S)=0$ and solenoidal $\bm B$ that obey Eq.~(\ref{constraint}) respectively.
Therefore under the conditions
\begin{eqnarray}&&\!\!\!\!\!\!\!\!\!\!\!\!\!\!
Re<{\tilde Re},\ \ Re_m<{\tilde Re}_m, \label{gla}
\end{eqnarray}
both $Q_i$ are negatively definite and the flow is globally stable.

The above considerations provide universal observation of the existence of threshold on $Re$ and $Re_m$ below which the flow is completely stable. Unfortunately it is hard to provide explicit form of this threshold as seen from the variational condition on $\bm u$ that provides maximum to $T/D$ under the constraints $\nabla\cdot\bm u=0$ and $\bm u(S)=0$ (we use with no ambiguity the same letters for $\bm u$ over which the maximum is sought
and $\bm u$ at which the maximum is attained). This reads
\begin{eqnarray}&&\!\!\!\!\!\!\!\!\!\!\!\!\!\!
D(\bm u) s \bm u+ T(\bm u)\nabla^2 \bm u \int_V\!\! \bm u s \bm u+D(\bm u)\nabla \lambda=0,
\end{eqnarray}
where the matrix $s$ is defined in Eq.~(\ref{soi1}) and $\lambda$ is the rescaled Lagrange multiplier associated with the solenoidality constraint. We find using $T(\bm u)/D(\bm u)={\tilde Re}^{-1}$ that \citep{serrin,drazin}
\begin{eqnarray}&&\!\!\!\!\!\!\!\!\!\!\!\!\!\!
s \bm u=-\nabla \lambda+\frac{1}{{\tilde Re}}\nabla^2\bm u,\ \ \nabla\cdot\bm u=0,\ \ \bm u(S)=0.  \label{varia}
\end{eqnarray}
The smallest ${\tilde Re}$ obeying the above equations provides the actual value of ${\tilde Re}$. Thus ${\tilde Re}$ is a generalized eigenvalue similar to that for usual Rayleigh quotients, see the description below and \cite{enc}.
We remark that since $T$ is a quadratic form then we can use instead of Eqs.~(\ref{dasd}) the equivalent definition ${\tilde Re}^{-1}=max_{\bm u}T(\bm u)$ where the maximum is taken under the constraints $\nabla\cdot\bm u=0$, $D(\bm u)=1$ and $\bm u(S)=0$. In this formulation ${\tilde Re}$ would arise from the Lagrange multiplier of the constraint $D(\bm u)=1$.

Eqs.~(\ref{varia}) resemble unsteady Stokes equations in frequency domain to which they reduce for $s$ which is proportional to the unit matrix (which is not a possible case since $tr (s)=0$ by incompressibility). They are sometimes, however not always, similar to those of linear stability analysis of $\bm v_0$, see \citep{drazin,serrin}. Unfortunately these equations are not solvable for general $s$ and even for simple $s$ are hard to solve \citep{drazin,serrin}. Thus no simple explicit expression for ${\tilde Re}$ is possible. Similar conclusion applies to
${\tilde Re}_m$. The way to proceed is by obtaining lower bounds for ${\tilde Re}$ and ${\tilde Re}_m$.

\subsection{Global stability in terms of strain tensor eigenvalues}\label{stabilitys}

Probably the simplest condition of global stability is found from the first line of Eq.~(\ref{xs}). 
We start from observing that Eq.~(\ref{st}) gives
\begin{eqnarray}&&\!\!\!\!\!\!\!\!\!\!\!\!\!\!
-\int_S (\bm v_0\cdot \bm B)(\bm {\hat n} \cdot  \bm B) dS= \int_{V'}\!\! B_iB_k\nabla_i v_k' d\bm x
\nonumber\\&&\!\!\!\!\!\!\!\!\!\!\!\!\!\!
\leq max[eig\left({s'}\right)]\int_{V'}\!\! B^2 d\bm x,
\end{eqnarray}
where we designated by $max[eig\left({s'}\right)]$ the maximal eigenvalue of ${s'}_{ik}=(\nabla_i {v'}_k+\nabla_k {v'}_i)/2$ over the domain $V'$ (here and below in many obvious cases the eigenvalues can be time-dependent). We can now pick in this inequality ${\bm v'}$ such that $max[eig\left({s'}\right)]$ is reached on the boundary $S$ where ${s'}$ coincides with $s$ (and as stated above ${\bm v'}(S)$ coincides with $\bm v_0(S)$). Such a field ${\bm v'}$ can be constructed since the only constraint in $V'$, which is $\nabla \cdot {\bm v'}=0$, is mild. For instance the representation ${\bm v'}=\nabla\times\bm A$ may be used, where the vector potential $\bm A$ has given first and second derivatives on $S$. We find then
\begin{eqnarray}&&\!\!\!\!\!\!\!\!\!\!\!\!\!\!
-\int_S (\bm v_0\cdot \bm B)(\bm {\hat n} \cdot  \bm B) dS\leq max_S[eig\left(s\right)]\int_{V'}\!\! B^2 d\bm x
\nonumber\\&&\!\!\!\!\!\!\!\!\!\!\!\!\!\!
\leq max_V[eig\left(s\right)]\int_{V'}\!\! B^2 d\bm x,
\end{eqnarray}
where the $max_S$ and $max_V$ designate maxima taken over $S$ and $V$ respectively. Then the first line of Eq.~(\ref{xs}) yields (unspecified integration domain stands for the whole space $V+V'$):
\begin{eqnarray}&&\!\!\!\!\!\!\!\!\!
\dot E\!\leq \!max_V[eig\left(s\right)]\int \!\! B^2 d\bm x\!-\!\eta\!\!\int_V (\nabla\!\times\!\bm B)^2d\bm x
\nonumber\\&&\!\!\!\!\!\!\!\!\!
\!+\!\left|min_V[eig\left(s\right)] \right|\int_V\!\! u^2 d\bm x\!-\!\nu\!\!\int_V \left(\nabla\!\times\! \bm u\right)^2 d\bm x,
\end{eqnarray}
where $min_V[eig\left(s\right)]$ is the minimum of the smallest eigenvalue of $s$ over $V$.
We assume that the flow is non-degenerate so that the incompressibility condition $tr (s)=0$ implies $max_V[eig\left(s\right)]>0$ and $min_V[eig\left(s\right)]<0$. Lower bounds on the Joule heating and kinetic energy dissipation in the volume $V$ are given by:
\begin{eqnarray}&&\!\!\!\!\!\!\!\!\!
\frac{\int_V \left(\nabla\times \bm u\right)^2 d\bm x}{\int_V u^2 d\bm x}\geq  \frac{x^2}{R^2},\ \ x\approx 4.493;\nonumber\\&&\!\!\!\!\!\!\!\!\!
\frac{\int_V \left(\nabla \times \bm B\right)^2d\bm x}{\int  B^2 d\bm x}\geq \frac{\pi^2}{R^2},\label{ins}
\end{eqnarray}
where $R$ is the smallest radius of the ball that encloses $V$. A detailed derivation of the above inequalities is given in Sections \ref{sec:fa} and \ref{sec:fb}. Moreover it is demonstrated that it is highly plausible that the equations above hold with $R=(4\pi/3V)^{2/3}$ which is the radius of the ball whose volume equals $V$ (inequalities in terms of smallest eigenvalues of certain linear operators are also possible, see \cite{backus} however they are implicit). We conclude from the last equations that
\begin{eqnarray}&&\!\!\!\!\!\!\!\!\!
\dot E\!\leq \!\left(max_V[eig\left(s\right)]-\frac{\eta \pi^2}{R^2}\right)\int \!\! B^2 d\bm x
\nonumber\\&& \!\!\!\!\!\!\!\!\!
\!+\!\left(\left|min_V[eig\left(s\right)] \right|-\frac{\nu x^2}{R^2}\right)\int_V\!\! u^2 d\bm x.
\label{conditions}
\end{eqnarray}
We conclude that universally, i.e. independent of the realisation of the perturbation velocity and magnetic field, the total energy of the perturbation decays provided
\begin{eqnarray}&&\!\!\!\!\!\!\!\!\!\!\!\!\!\!
\frac{max_V[eig\left(s\right)] R^2}{\eta\pi^2}\leq 1;\ \ \frac{\left|min_V[eig\left(s\right)] \right|R^2}{\nu x^2}\leq 1. \label{crt}
\end{eqnarray}
Thus decay takes place provided the dissipation coefficients $\eta$ and $\nu$ exceed a critical value. The left-hand sides of the inequalities above could be considered as definitions of the corresponding magnetic and kinetic Reynolds numbers, cf. \cite{serrin}. For that purpose $V_0$ is expressed in terms of the  maximal and minimal eigenvalues of $s$, respectively, while $L$ is identified with $R$. The inequality on $\eta$ generalizes that of \cite{backus} who studied linear magnetic stability problem in a given flow $\bm v_0$ and assumed that on the boundary $\bm v$ either vanishes or is a rigid rotation. The inequality on $\nu$ generalizes that of \cite{serrin} who studied hydrodynamic stability without the magnetic field and used a combination alternative to $x^2/R^2$, see below. The last reference also contains an example of application to Couette flow, cf. below. For single-scale flow both the maximum and minimum eigenvalues are of order $V_0/L$. Multi-scale flow, such as developed turbulence is considered later.

\subsection{Global stability in terms of the kinetic and magnetic Reynolds numbers} \label{glo}

In this section we provide another stability criterion which is given via quantities that resemble the usual definition of the Reynolds numbers. We rewrite Eq.~(\ref{xs}) using the identities $(\bm B\cdot \nabla)\bm B=\nabla B^2/2+\bm j\times \bm B$ and $(\bm u\cdot \nabla)\bm u=\nabla u^2/2+\bm w\times \bm u$ as
\begin{eqnarray}&&\!\!\!\!\!\!\!\!\!\!\!\!\!\!
\frac{dE}{dt}\!=\!-\int_V\!\! \left(\bm j \!\cdot\!(\bm B\!\times\! \bm v_0)\!+\!\bm w\!\cdot\!(\bm v_0\!\times\! \bm u) \!+\!\nu w^2\!+\!\eta j^2\right) d\bm x, \label{fedor}
\end{eqnarray}
where we used the divergence theorem, $\nabla\cdot\bm v_0=0$ and the condition of vanishing normal velocity on $S$. We observe that for any positive $\kappa$ and ${\tilde \kappa}$
\begin{eqnarray}&&\!\!\!\!\!\!\!\!\!\!\!\!\!\!
\pm 2\bm w\cdot(\bm v_0\times \bm u)\leq \kappa w^2+\kappa^{-1}(\bm v_0\times \bm u)^2;\nonumber\\&&\!\!\!\!\!\!\!\!\!\!\!\!\!\!
\pm 2\bm j \cdot(\bm B\times \bm v_0)\leq {\tilde \kappa} j^2+{\tilde \kappa}^{-1}(\bm B\times \bm v_0)^2,
\label{inequality}
\end{eqnarray}
cf. \cite{serrin} who used a similar inequality which fits less in our case.
We find
\begin{eqnarray}&&\!\!\!\!\!\!\!\!\!\!\!\!\!\!
\frac{dE}{dt}
\!\leq\! \int_V\!\! \left(\frac{(\bm v_0\!\times\! \bm u)^2}{2\kappa}\!+\!\frac{\kappa w^2}{2}\!-\!\nu w^2
\right.\nonumber\\&&\!\!\!\!\!\!\!\!\!\!\!\!\!\!\left.
+\frac{(\bm B\!\times \!\bm v_0)^2}{2{\tilde \kappa}}\!+\!\frac{{\tilde \kappa}j^2}{2}\!-\!\eta j^2\right)d\bm x , \label{kappa}
\end{eqnarray}
The values of $\kappa$ and ${\tilde \kappa}$ that minimize the last line are
\begin{eqnarray}&&\!\!\!\!\!\!\!\!\!\!\!\!\!\!
\kappa^*\!=\!\left(\!\frac{\int_V \!(\bm v_0\!\times\! \bm u)^2 d\bm x}{\int_V w^2d\bm x}\!\right)^{1/2},\ {\tilde \kappa}^*\!=\!\left(\!\frac{\int_V\!(\bm B\!\times\! \bm v_0)^2 d\bm x}{\int_V j^2d\bm x}\!\right)^{1/2}.
\end{eqnarray}
We find using these values
\begin{eqnarray}&&\!\!\!\!\!\!\!\!\!\!\!\!\!
\dot E\!\leq \!\sqrt{\int_V\!\! w^2d\bm x\int_V\!\! (\bm v_0\!\times\! \bm u)^2 d\bm x }\left(1\!-\!\nu\sqrt{\frac{\int_V\!w^2d\bm x}{\int_V\! (\bm v_0\!\times\! \bm u)^2 d\bm x}}\right)
\nonumber\\&&\!\!\!\!\!\!\!\!\!\!\!\!\!
+\!\sqrt{\int_V\!\! j^2d\bm x\int_V\!\! (\bm B\!\times\! \bm v_0)^2 d\bm x }\left(1\!-\!\eta\sqrt{\frac{\int_V\! j^2 d\bm x}{\int_V (\bm B\!\times\! \bm v_0)^2 d\bm x }}\right).\label{sq}
\end{eqnarray}
We start the study of this inequality from its hydrodynamic part in the first line. We see that it predicts universal decay of energy for purely hydrodynamic flow for supercritical viscosity described by
\begin{eqnarray}&&\!\!\!\!\!\!\!\!\!\!\!\!\!\!
\nu^2 \min_{\bm u}\left(\frac{\int_V \left(\nabla \bm u\right)^2d\bm x}{\int_V u^2v_0^2 d\bm x}\right)>1;\ \ \ \frac{dE_{kin}}{dt}<0. \label{stp}
\end{eqnarray}
This stability criterion coincides with that derived in \cite{serrin} who used in Eq.~(\ref{kappa}) the value $\kappa=\nu$ producing
\begin{eqnarray}&&\!\!\!\!\!\!\!\!\!\!\!\!
\frac{dE_{kin}}{dt}\leq \int_V\!\! \left(\frac{u^2v_0^2}{2\nu}-\frac{\nu}{2}\left(\nabla \bm u\right)^2\right)d\bm x
\nonumber\\&&\!\!\!\!\!\!\!\!\!\!\!\!
=\int_V\!\! \frac{u^2v_0^2}{2\nu}d\bm x\left(1-\nu^2\frac{\int_V \left(\nabla \bm u\right)^2d\bm x}{\int_V u^2v_0^2 d\bm x}\right),
\end{eqnarray}
which is universally negative under the same condition as that provided in Eq.~(\ref{st}). The difference between the first line of Eq.~(\ref{sq}) and the above equation is only in the magnitude of the derivative, the signs coincide.

It is still difficult to produce a universal criterion using Eq.~(\ref{stp}) since the variational equations are not solvable for general $\bm v_0$. Therefore further loosening of the bound is made by employing the inequality
$\int_V (\bm v_0\times \bm u)^2 d\bm x\leq v_{0max}^2\int_V u^2 d\bm x$ where $v_{0max}$ is the maximal value of $v_0$ over $V$. Similarly we use $\int_V (\bm B\times \bm v_0)^2 d\bm x \leq v_{0max}^2\int_V B^2 d\bm x$.
We find that
\begin{eqnarray}&&\!\!\!\!\!\!\!\!
\dot E\!\leq\! \sqrt{\int_V \!\!w^2d\bm x\int_V \!\!(\bm v_0\!\times \!\bm u)^2 d\bm x }\left(1\!-\!\sqrt{\frac{\nu^2\int_V\left(\nabla\times \bm u\right)^2d\bm x}{v_{0max}^2\int_V u^2 d\bm x}}\right)
\nonumber\\&&\!\!\!\!\!\!\!\!
+\sqrt{\!\int_V \!\!j^2d\bm x \!\int_V \!\!(\bm B\!\times\! \bm v_0)^2 d\bm x }\left(\!1\!-\!\sqrt{\frac{\eta^2\int_V \!\!\left(\nabla\!\times \!\bm B\right)^2 d\bm x}{v_{0max}^2\int_V B^2 d\bm x }}\right).\nonumber
\end{eqnarray}
Employing the following bound (see detailed derivation in Section \ref{sec:fc}):
\begin{equation}
\frac{\int_V \left(\nabla \times \bm B\right)^2d\bm x}{\int_V  B^2 d\bm x}\geq \frac{x_1^2}{R^2},\ \  x_1\approx 3.506.  \label{re0}
\end{equation}
as well as the first of Eqs.~(\ref{ins}) we conclude that $\dot E<0$ independently of the structure of the perturbation if
\begin{eqnarray}&&\!\!\!\!\!\!\!\!\!\!\!\!
Re\equiv \frac{v_{0max} R}{\nu }<x\approx 4.493,\nonumber\\&&\!\!\!\!\!\!\!\!\!\!\!\!
Re_m\equiv \frac{v_{0max} R}{\eta}<x_1\approx 3.506.
\label{criteria}
\end{eqnarray}
Here $R$ is the radius of the smallest ball enclosing the domain. Under the assumption of validity of generalized Rayleigh conjecture, smaller $R$ given by $(3V/4\pi)^{1/3}$ can be used. The above equations provide the stability in terms of $Re$ and $Re_m$ whose definition is one
of the possible usual definitions of the Reynolds numbers. For purely hydrodynamic flow $v_{0max} (3V/4\pi)^{1/3}<\nu x$ is a tighter stability bound than in \cite{serrin}.

\section{Variational inequalities}\label{variational}
In this section we prove and discuss the necessary inequalities that have been used in order to derive the stability conditions so far.
\subsection{Generalized Rayleigh-Faber-Krahn inequality for minimal dissipation} \label{sec:fa}

In this section we provide a proof for the first inequality in Eq. (\ref{ins}), namely we seek $\lambda_1$ which is the largest number for which
\begin{eqnarray}&&\!\!\!\!\!\!\!\!\!\!\!\!\!\!
\frac{\int_V \left(\nabla \bm u\right)^2d\bm x}{\int_V u^2 d\bm x}\geq \lambda_1,
\label{lambda1}
\end{eqnarray}
holds for any solenoidal $\bm u$ vanishing on $S$. This rather fundamental problem is intractable analytically since $\lambda _1$ depends not only on the domain's volume $V$ but also on its shape. In order to overcome this hurdle, a less restrictive question may be posed which is: for a given volume $V$ what shape yields the smallest $\lambda _1$ (which may now be unambiguously denoted by $\lambda _1(V)$).

The route to answering that alternative problem starts with Rayleigh, who considered the minimum of the quotient
\begin{equation}
\frac{\int_V (\nabla \phi)^2d\bm x}{\int_V \phi^2 d\bm x}.
\end{equation}
over all scalar functions $\phi (\bm x)$ that vanish at the boundary.  The minimal value, denoted by $\lambda_1 ^L$ is the smallest Dirichlet eigenvalue of the Laplacian operator on $V$, i.e. the smallest positive solution of $\nabla ^2 \phi=-\lambda _1^L \phi$ with the Dirichlet zero boundary conditions. 


Rayleigh conjectured that for domains with given volume $V$ the minimal $\lambda^L_1$ is attained for a ball (the original formulation was for two-dimensional case of a membrane). In retrospect this seems to be inevitable because of the isotropy of the problem. For a ball of radius $R$ the eigenfunction with smallest eigenvalue is given by is the spherically symmetric solution $\sin(\pi r/R)/r$ of the Laplace equation that does not have zeros inside the domain. The eigenvalue is $\pi^2/R^2$ and writing $R^3=3V/(4\pi)$ we find that the conjecture is $\lambda^L_1\geq \pi^2 (4\pi/3V)^{2/3}$. Rayleigh's conjecture has been proven and is known as Rayleigh-Faber-Krahn inequality, see details in \cite{enc}.

Moving on, we notice that if the solenoidality requirement on ${\bm u}$ is relaxed, $\lambda _1$ in Eq. (\ref{lambda1}) may be inferred by the Rayleigh-Faber-Krahn inequality. We find
\begin{eqnarray}&&\!\!\!\!\!\!\!\!\!\!\!\!\!\!
\int_V (\nabla u_i)^2d\bm x\geq \pi^2\left(\frac{4\pi}{3V}\right)^{2/3}\int_V u_i^2 d\bm x;\nonumber\\&&\!\!\!\!\!\!\!\!\!\!\!\!\!\!
\int_V (\nabla \bm u)^2d\bm x\geq \pi^2\left(\frac{4\pi}{3V}\right)^{2/3}\int_V u^2 d\bm x. \label{krahn}
\end{eqnarray}
where the last inequality is obtained by summing the first over $i$.

Obviously, we have not yet achieved our goal, as inequality (\ref{krahn}) holds for any $\bm u$ that satisfies $\bm u(S)=0$ but not necessarily the solenoidality condition. We therefore need to modify (\ref{krahn}) by taking the constraint $\nabla \cdot {\bm u}=0$ into account. Thus, employing a Lagrange multiplier in order to ensure that, the variational formulation of (\ref{lambda1}) yields the following equation:
\begin{eqnarray}&&\!\!\!\!\!\!\!\!\!\!\!\!\!\!
-\lambda \bm u=-\nabla p+\nabla^2 \bm u,\ \ \nabla\cdot\bm u=0, \ \ \bm u(S)=0, \label{op}
\end{eqnarray}
where $p$ is the rescaled Lagrange multiplier ensuring the solenoidality constraint. The value of $\lambda_1$ is then given by the smallest positive solution of the above eigenvalue equation for $\lambda$.

{\bf Generalized Rayleigh conjecture}---
Since Eqs. (\ref{op}) do not break isotropy, it seems inevitable that the Rayleigh conjecture holds for them as well. This means that the minimal $\lambda_1$ over all domains with given volume $V$ is attained for a ball of radius $(3V/4\pi)^{1/3}$. Thus we consider the solution of the above equations for ball of radius $R$.

A detailed solution of the eigenvalue problem (\ref{op}) is given in Appendix \ref{appendix:A}. Here we sketch the solution method and state the main relevant results. The eigenfunctions are obtained as linear combinations of vector spherical harmonics (VSH) labeled by indices $l$ and $m$ similar to those of the usual spherical harmonics. For each given $l$ and $m$ there are two linearly independent modes, as fixed by the number of components of the vector (three) minus one due to incompressibility constraint. For one of the modes the pressure is zero and the solution is given by Eq.~(\ref{modes}). To get another mode we observe by taking the divergence of Eq.~(\ref{op}) that $\nabla^2 p=0$.  This means that the Lagrange multuplier $p$ can be written as a superposition of spherical harmonics that are regular in the origin, i.e. $r^lY_{lm}(\theta, \phi)$. We look therefore for solutions $\bm u_{lm}$ of Eq.~(\ref{op}) that correspond to $p=r^lY_{lm}(\theta, \phi)$ and satisfy
\begin{eqnarray}&&\!\!\!\!\!\!\!\!\!\!\!\!\!\!\!\!\!
\lambda_{lm}\bm u_{lm} \!+\!\nabla^2 \bm u_{lm}\!=\!\nabla \left(r^lY_{lm}\right)\!. \label{qw}
\end{eqnarray}
The solution of this equation gives the remaining mode given by Eq.~(\ref{fomrs}). The smallest eigenvalue is attained at the mode with zero pressure and $l=m=1$. It is given by $x^2/R^2$ where $x$ solves the equation $x=\tan x$. It is found that using the solenoidality condition we can tighten the bound given by Eq.~(\ref{krahn}) as
\begin{eqnarray}&&\!\!\!\!\!\!\!\!\!\!\!\!\!\!
\frac{\int_V (\nabla \bm u)^2d\bm x}{\int_V u^2 d\bm x}\geq x^2\left(\frac{4\pi}{3V}\right)^{2/3},\ \ x=\tan x \approx 4.493, \label{kr1}
\end{eqnarray}
where we expressed the ball radius $R$ via the ball's volume $V$. This concludes the proof of the first inequality in (\ref{ins}). Stricktly speaking, this proof is valid only for a ball while for any other arbitrary shape it hinges on the generalised Rayleigh conjecture. The proof of the latter is beyond the scope of the current work.

\subsection{Connection to freely decaying flow}
Assuming the validity of the generalised Rayleigh conjecture, we have derived above in passing what seems as a rather fundamental fluid mechanical result, namely we have derived the rate of the logarithmic change of the kinetic energy in freely decaying incompressible flow.  In order to see that we consider the case of $v_0=\bm B=0$, for which the logarithmic derivative of the kinetic energy (now the true kinetic energy) is given by:
\begin{equation}
-\frac{d\ln E_{kin}}{dt} =\frac{\int_V \left(\nabla \bm u\right)^2d\bm x}{\int_V u^2 d\bm x}. \label{lsd1}
\end{equation}
Combining that with eq. (\ref{kr1}) we obtain:
\begin{eqnarray}&&\!\!\!\!\!\!\!\!\!\!\!\!\!\!
-\frac{d\ln E_{kin}}{dt}\geq  \nu x^2\left(\frac{4\pi}{3V}\right)^{2/3},\ \ x=\tan x \approx 4.493,
\end{eqnarray}
where it can be realised as equality only if the domain is a ball, otherwise the inequality is strict.

Further detailed information may be obtained for a spherical domain from the eigenmodes described above, that actually provide solutions to the problem of freely decaying unsteady Stokes flow in a ball. Such a flow obeys
\begin{eqnarray}&&\!\!\!\!\!\!\!\!\!\!\!\!\!\!
\partial_t\bm u=-\nabla p+\nu\nabla^2\bm u,\ \ \nabla\cdot \bm u=0,\ \ \bm u(x=R)=0.\label{fre}
\end{eqnarray}
The solutions are the zero pressure modes
\begin{eqnarray}&&\!\!\!\!\!\!\!\!\!\!\!\!\!\!
\bm u_{lmn}\!=\!e^{-\nu\lambda_{ln}t}j_{l}(\sqrt{\lambda_{ln}} r)\bm \Phi_{lm},\ \ j_{l}(\sqrt{\lambda_{ln}} R)\!=\!0,
\label{zerop}
\end{eqnarray}
as well as modes with finite pressure
\begin{eqnarray}
p_{lmn}&=&\nu\exp\left(-\nu \lambda^p_{ln}t\right)r^lY_{lm}(\theta, \phi),\\
\bm u_{lmn}&=&
\exp\left(-\nu \lambda^p_{ln}t\right)\left(\frac{{\tilde c}^{r}_{ln} j_{l}(\sqrt{\lambda_l} r)}{r}\!+\!\frac{lr^{l-1}}{\lambda_l}\right)\bm Y_{lm}\nonumber \\&&\!\!\!\!\!\!\!\!\!\!\!\!
+\exp\left(-\nu \lambda^p_{ln}t\right)\left(\frac{{\tilde c}^{r}_{ln}\left(rj_{l}(\sqrt{\lambda_{ln}} r)\right)' }{l(l+1)r}
\!+\!\frac{r^{l-1}}{\lambda_{ln}}\right)\bm \Psi_{lm}, \nonumber
\end{eqnarray}
where $\bm u_{lmn}(r=R)=0$ and $\bm \Phi_{lm}$, $\bm Y_{lm}$, $\bm \Psi_{lm}$ are vector spherical harmonics defined in Appendix \ref{appendix:A}. Solution of Eq.~(\ref{fre}) with arbitrary initial conditions can be obtained by the linear superposition of the above two sets of modes. The asymptotic behavior of the solution at large times is determined by the slowest decaying mode (assuming that it is present in the initial conditions). The slowest mode has zero pressure and velocity obeying
\begin{eqnarray}&&\!\!\!\!\!\!\!\!\!\!\!\!\!\!
\bm u(t)\!\propto\! \exp\left(-\frac{x^2\nu t}{R^2}\right)\left[\frac{R}{xr}\sin\left(\frac{x r}{R}\right)-\cos\left(\frac{x r}{R}\right)\right]\frac{\bm \Phi_{lm}}{r},\nonumber
\end{eqnarray}
where $x$ is defined in Eq.~(\ref{kr1}). Since this mode has zero pressure then the pressure decay is determined by a larger eigenvalue and is faster. It is given by
\begin{eqnarray}&&\!\!\!\!\!\!\!\!\!\!\!\!\!\!
p(t)\propto \exp\left(-\frac{{\tilde x}^2\nu t}{R^2}\right)rY_{1m}(\theta, \phi),
\end{eqnarray}
where ${\tilde x}$ is defined by
\begin{eqnarray}&&\!\!\!\!\!\!\!\!\!\!\!\!\!\!
3\sin {\tilde x}-3{\tilde x}\cos {\tilde x}-{\tilde x}^2\sin {\tilde x}=0,\ \ {\tilde x}\approx 5.763,
\end{eqnarray}
as can be seen from the formulas in Appendix \ref{appendix:A}.

It is quite remarkable that the slowest decaying mode has zero pressure (here the arbitrary constant in the pressure is set to zero). This result can be compared with the well-known decay problem in unbounded fluid. There Eq.~(\ref{fre}) can be solved by the Fourier transform which demonstrates that the incompressible transversal modes of the fluid decouple from the pressure \cite{reichl}. The slowest decaying mode in a ball has the same property of decoupling from the pressure. It is of much interest to see if the decoupling generalizes to arbitrary shape of the container.

\subsection{Universal decay of magnetic component of energy}\label{sec:fb}

While proving in the previous section the first inequality in (\ref{ins}) and exploring its physical consequences, we focus in this section on the second inequality in (\ref{ins}) in order to complete the proof of the stability conditions (\ref{conditions}).  In analogy with eq. (\ref{lsd1}), the left hand side of the second inequality in (\ref{ins}) may be expressed as the logarithmic rate of the magnetic energy dissipation in a stationary fluid.  Thus, we seek the lower bound on the latter over all admissible $\bm B$ fields:
\begin{eqnarray}&&\!\!\!\!\!\!\!\!\!\!\!\!\!\!
-\frac{1}{\eta}\frac{d\ln E_m}{dt} =\frac{\int_V \left(\nabla \times \bm B\right)^2d\bm x}{\int_{V+V'} B^2 d\bm x}\geq \lambda,\nonumber\\&&\!\!\!\!\!\!\!\!\!\!\!\!\!\!
\lambda\equiv \min_{\bm B}\frac{\int_V \left(\nabla \times \bm B\right)^2d\bm x}{\int_{V+V'} B^2 d\bm x}, \label{fe}
\end{eqnarray}
For writing down the variational condition on $\bm B$ that determines $\lambda$ we observe that (recalling that $\bm j=\nabla \times \bm B$ and $\nabla \times \bm j=-\nabla^2\bm B$)
\begin{eqnarray}&&\!\!\!\!\!\!\!\!\!\!\!\!\!\!
\delta\frac{\int_V\!\! \left(\nabla\! \times \!\bm B\right)^2d\bm x}{2}\!=\!\int_V\!\! \bm j\!\cdot\!\left(\nabla\! \times \!\delta \bm B\right)d\bm x
\nonumber\\&&\!\!\!\!\!\!\!\!\!\!\!\!\!\!
=\int_S \!\!\delta \bm B \!\cdot\! \left(\bm j\!\times \! \bm {\hat n}\right)
 dS\!-\!\int_V\!\! \delta \bm B  \!\cdot\!\nabla^2\bm B d\bm x,
\end{eqnarray}
where $\bm j$ on $S$ is defined by continuation from inside the domain and we used $\epsilon_{ikl}j_i\nabla_k\delta B_l=\epsilon_{ikl}\nabla_k (j_i\delta B_l)-\epsilon_{ikl}\delta B_l \nabla_k j_i$.
The variation of $\bm B(S)$ is constrained by Eq.~(\ref{constraint}). Employing the latter we find:
\begin{eqnarray}&&\!\!\!\!\!\!\!\!\!
\int_S \!\!\left(\bm j\!\times \! \bm {\hat n}\right)
\!\cdot\! \delta \bm B dS\!=\!-\!\!\int_S\!\! dS\left(\bm j\!\times  \!\bm {\hat n}\right)\!\cdot\! \nabla\!\! \int_S \!\!G(\bm x, \bm x')\delta B_n(\bm x')dS'
\nonumber\\&&\!\!\!\!\!\!\!\!\!
=-\int_S\!\! dS \bm j\!\cdot\! \left(\bm {\hat n}\!\times\!\nabla\right)  \int_S \!\!G(\bm x, \bm x')\delta B_n(\bm x')dS',
\end{eqnarray}
where the variation $\delta B_n$ is already unconstrained, $\bm x$, $\bm x'$ refer to $S$ and $S'$ respectively and $G(\bm x, \bm x')$ on the surface is found by approaching $S$ from outside the fluid domain. The formula is well-defined since it only involves tangential derivatives of $G$. We have using that $\int_S {\hat n}\times \nabla f(\bm x) dS=0$ for any function $f$ that
\begin{eqnarray}&&\!\!\!\!\!\!\!\!\!
\int_S \!\!\left(\bm j\!\times \! \bm {\hat n}\right)
\!\cdot\! \delta \bm B dS\!=\!\int_S \!\!G(\bm x, \bm x')\delta B_n(\bm x')dS' \int_S\!\!  \left(\bm {\hat n}\!\times\!\nabla\right)\!\cdot \!\bm jdS .\nonumber
\end{eqnarray}
We find using $\left(\bm {\hat n}\!\times\!\nabla\right)\cdot \bm j =\bm {\hat n} \cdot (\nabla\times \bm j)=-\bm {\hat n} \cdot \nabla^2\bm B$ and combining the equations that
\begin{eqnarray}&&\!\!\!\!\!\!\!\!\!
\delta\int_V\!\! \left(\nabla\! \times \!\bm B\right)^2d\bm x\!=\!-2\!\int_V\!\! \delta \bm B  \!\cdot\!\nabla^2\bm B d\bm x
\nonumber\\&&\!\!\!\!\!\!\!\!\!
-2\int_S \!\!G(\bm x, \bm x')\delta B_n(\bm x')dS' \int_S\!\! dS \bm {\hat n} \cdot \nabla^2\bm B.
\end{eqnarray}
We obtain using this equation and considering the variation of $\int  B^2 d\bm x=\int_{V}  B^2 d\bm x+\int_{V'}  B^2 d\bm x$ that
\begin{eqnarray}&&\!\!\!\!\!\!\!\!\!
\delta \frac{\int_V \!\!\left(\nabla \!\times \!\bm B\right)^2d\bm x}{2\int_{V+V'}\!\!  B^2 d\bm x}\!=\!\frac{(-1)}{\int_{V+V'}\!\!  B^2 d\bm x}\left(\!
\int_V\!\! \delta \bm B  \!\cdot\!\left(\nabla^2\bm B\!+\!\lambda\bm B\right)d\bm x
\right.\nonumber\\&&\!\!\!\!\!\!\!\!\!\left.
+\int_S \!\!G(\bm x, \bm x')\delta B_n(\bm x')dS' \int_S\!\! dS \bm {\hat n} \!\cdot\!\left(\nabla^2\bm B\!+\!\lambda\bm B\right)\right),
\end{eqnarray}
where we used for $E_e=\int_{V'}B^2d\bm x/2$ the form given by Eq.~(\ref{defe}). The corresponding variation for the Lagrange multiplier ${\tilde p}$ ensuring the solenoidality is
\begin{eqnarray}&&\!\!\!\!\!\!\!\!\!\!\!\!\!\!
\delta \int_{V} {\tilde p}\nabla\cdot\bm Bd\bm x=-\int_{V} \delta\bm B\cdot \nabla {\tilde p} d\bm x+\int_S {\tilde p} \delta B_n dS.
\end{eqnarray}
We find passing to rescaled multiplier $p={\tilde p} \int  B^2 d\bm x$ that the variational condition on $\bm B$ is $\nabla\cdot\bm B=0$ and
\begin{eqnarray}&&\!\!\!\!\!\!\!\!\!\!\!\!\!\!
\nabla^2\bm B\!=\!\nabla p\!-\!\lambda\bm B,\ \ p(S)\!=\! -\!\int_S \!\!G(\bm x, \bm x') \bm {\hat n}\!\cdot\!\nabla p(\bm x')  dS'. \label{sa}
\end{eqnarray}
These must hold in $V$ and on $S$ and where we used the symmetry of the restriction of the Green's function to $S$, see the remark after Eq.~(\ref{ref}). We used that the limiting values of $\nabla^2\bm B+\lambda\bm B$ on $S$, approached from inside the domain, coincide with the similar values of $\nabla p$. The magnetic field also obeys Eq.~(\ref{constraint}) ensuring that $\bm B$ can be continuously extended from $V$ to $V'$ as a potential field.

Applying the divergence operator to both sides of the first of Eqs.~(\ref{sa}) results in $\nabla ^2 p=0$. Together with the homogeneous integral Robin-type boundary condition it implies that the only solution for $p$ is $p=0$, see a similar conclusion in e.g. \cite{sb}. This result may be proven formally by observing that $-\int_S G(\bm x, \bm x') \bm {\hat n}\cdot\nabla p(\bm x')  dS'$ defines a harmonic function $p_o(\bm x)$ for $\bm x$ in $V'$. Then applying expansion in the spherical harmonics to the harmonic functions $p$ in $V$ and $p_o$ in $V'$ it is readily seen that the boundary condition $p_o(S)=p(S)$ implies $p=p_o=0$.

We conclude that the variational problem reduces to finding the smallest positive solution $\lambda$ of
\begin{eqnarray}&&\!\!\!\!\!\!\!\!\!\!\!\!\!\!
\nabla^2\bm B=-\lambda\bm B,\ \ \bm x\ in\ V; \ \ \bm B=\nabla\phi,\ \ \bm x\ outside\ V,
\end{eqnarray}
which is continuous on $S$ and where $\nabla\cdot\bm B=0$ is imposed.

Considering once again a spherical domain of radius $R$, and using the definition of the VSH in Eq.~(\ref{vsh}) the general form of the solution in $V'$ is
\begin{eqnarray}&&\!\!\!\!\!\!\!\!\!\!\!\!\!\!
\bm B=\nabla \sum_{l=1}^{\infty}\sum_{m=-l}^l \frac{b_{lm}Y_{lm}(\theta, \phi)}{r^{l+1}}
\nonumber\\&&\!\!\!\!\!\!\!\!\!\!\!\!\!\!
=\sum_{l=1}^{\infty}\sum_{m=-l}^l \frac{b_{lm}\left(\bm \Psi_{lm}-(l+1)\bm Y_{lm}\right)}{r^{l+2}},\ \ \bm x\ in\ V',\label{dw}
\end{eqnarray}
with certain constant coefficients $b_{lm}$ (the requirement that $r^3B$ is bounded does not allow $l=0$ term in the series above). Similar expansion for $\bm B$ inside $V$ demonstrates that the
the set of eigenmodes provided by Eq.~(\ref{modes}) can only fit $\bm B(V')$ in Eq.~(\ref{dw}) if $\bm B(V')$ is trivial. Thus the problem reduces to the previously studied problem on $\bm u$
in $V$ however with zero pressure. The velocity modes with zero pressure provide immediately also the corresponding mode of $\bm B$ that vanishes in $V'$ and in $V$ reads
\begin{eqnarray}&&\!\!\!\!\!\!\!\!\!\!\!\!\!\!
\bm B\!=\!\bm B^{(2)}_{lmn}\!=\!j_{l}(\sqrt{\lambda_{ln}} r)\bm \Phi_{lm},\ \ j_{l}(\sqrt{\lambda_{ln}} R)\!=\!0, \label{pres}
\end{eqnarray}
cf. Eq.~(\ref{zerop}). The smallest eigenvalue is given by Eq.~(\ref{fx}). The other set of eigenmodes is found by removing in Eq.~(\ref{fomrs}) the contribution of terms due to pressure. We find in $V$
\begin{eqnarray}&&\!\!\!\!\!\!\!\!\!\!\!\!\!\!\!
\bm B=\bm B_{lmn}\!=\!\frac{ j_{l}(\sqrt{\lambda_{ln}} r)}{r}\bm Y_{lm}
\!+\!\frac{\left(rj_{l}(\sqrt{\lambda_{ln}} r)\right)' }{l(l+1)r}\bm \Psi_{lm}, \label{mode}
\end{eqnarray}
where we set the multiplicative constant ${\tilde c}^{r}_{ln}$ to one. The condition that this solution can agree with Eq.~(\ref{dw}) on $S$ yields
\begin{eqnarray}&&\!\!\!\!\!\!\!\!\!\!\!\!\!\!\!
\frac{\left(rj_{l}(\sqrt{\lambda_{ln}} r)\right)' }{l(l+1)j_{l}(\sqrt{\lambda_{ln}} r)}|_{r=R}=-\frac{1}{l+1};\ \
\frac{j'_{l}(x_{ln}) }{j_{l}(x_{ln})}=-\frac{l+1}{x_{ln}},\ \
\label{eig}
\end{eqnarray}
where we introduced $x_{ln}\equiv\sqrt{\lambda_{ln}} R$. The smallest eigenvalue holds for $l=n=1$ that obeys
$x_{11}^2\sin x_{11}=0$. The smallest non-trivial solution is $x_{11}=\pi$ which is smaller than Eq.~(\ref{fx}). We conclude that
\begin{eqnarray}&&\!\!\!\!\!\!\!\!\!\!\!\!
\frac{\int_V \left(\nabla \times \bm B\right)^2d\bm x}{\int  B^2 d\bm x}\geq \frac{\pi^2}{R^2}. \label{re}
\end{eqnarray}
This concludes the proof of the second inequality in (\ref{ins}) for a spherical domain. Extending this result to other shapes is plausible but requires a proof that is beyond the scope of the current work. This inequality was obtained in \cite{backus} by another method. Similarly to the previous section the modes obtained above provide the eigenmodes for the problem of freely decaying magnetic field, cf. the previous section.

\subsection{Further inequality}\label{sec:fc}
In this section we seek the minimum of $\int_V \left(\nabla \times \bm B\right)^2d\bm x/\int_V B^2 d\bm x$ whose difference from Eq.~(\ref{fe}) is that the denominator includes the magnetic energy inside $V$ and not in the whole space. Expression for that quantity is needed in order to derive the stability criteria given in Eq. (\ref{criteria}). The variational equations in this case are readily seen to be given by Eq.~(\ref{sa}) with the only change that the boundary conditions on $p$ become $p(S)=p_o(S)$ with $p_o(\bm x)$ defined by
\begin{eqnarray}&&\!\!\!\!\!\!\!\!\!\!\!\!\!\!
p_o(\bm x)\!\equiv\! -\int_S \!\!G(\bm x, \bm x') \bm {\hat n}\!\cdot\!\left(\nabla p(\bm x')\!-\!\lambda\bm B(\bm x')\right)  dS'
\nonumber\\&&\!\!\!\!\!\!\!\!\!\!\!\!\!\!
\!=\!-\lambda \phi(\bm x)\!-\!\int_S \!\!G(\bm x, \bm x') \bm {\hat n}\!\cdot\!\nabla p(\bm x') dS',
\end{eqnarray}
where we used Eq.~(\ref{ref}) as the definition of $\phi$. We observe that $p_o$ is harmonic in $V'$ satisfying
\begin{eqnarray}&&\!\!\!\!\!\!\!\!\!\!\!\!\!\!
\nabla^2p_o=0,\ \ \bm {\hat n}\!\cdot\!\nabla p_o(S)=\bm {\hat n}\!\cdot\!\nabla p(S)-\lambda B_n(S). \label{bs}
\end{eqnarray}
We find immediately that the eigenmodes in Eq.~(\ref{pres}) are also eigenmodes of the problem at hand. The other set of eigenmodes $\bm B_{lm}$ is found by looking for solutions for $p=r^lY_{lm}(\theta, \phi)$. We find using the results of the previous section that
\begin{eqnarray}&&\!\!\!\!\!\!\!\!\!\!\!\!\!\!\!
\bm B_{lm}\!=\!\left(\frac{{\tilde c}^{r}_{lm} j_{l}(\sqrt{\lambda_l} r)}{r}\!+\!\frac{lr^{l-1}}{\lambda_l}\right)\bm Y_{lm}
\nonumber\\&&\!\!\!\!\!\!\!\!\!\!\!\!\!\!\!
+\left(\frac{{\tilde c}^{r}_{lm}\left(rj_{l}(\sqrt{\lambda_l} r)\right)' }{l(l+1)r}
\!+\!\frac{r^{l-1}}{\lambda_l}\right)\bm \Psi_{lm},\label{b}
\end{eqnarray}
which yields the following normal component on the surface
\begin{eqnarray}&&\!\!\!\!\!\!\!\!\!\!\!\!\!\!\!
\bm {\hat n}\!\cdot\!\bm B_{lm}\!=\!\left(\frac{{\tilde c}^{r}_{lm} j_{l}(\sqrt{\lambda_l} R)}{R}\!+\!\frac{lR^{l-1}}{\lambda_l}\right) Y_{lm}(\theta, \phi). 
\end{eqnarray}
The potential corresponding to the Neumann boundary condition provided by the above equation is given by:
\begin{eqnarray}&&\!\!\!\!\!\!\!\!\!\!\!\!\!\!\!
\phi=-\left(\frac{{\tilde c}^{r}_{lm} j_{l}(\sqrt{\lambda_l} R)}{R}\!+\!\frac{lR^{l-1}}{\lambda_l}\right)\frac{R^{l+2}Y_{lm}(\theta, \phi)}{(l+1)r^{l+1}}.
\end{eqnarray}
Requiring now that tangential components of the magnetic field $\nabla\phi$ on the boundary $S$ match those obtained from approaching the surface from $V$, as described by Eq.~(\ref{b}), we obtain:
\begin{eqnarray}&&\!\!\!\!\!\!\!\!\!\!\!\!\!\!\!
\frac{\lambda_l{\tilde c}^{r}_{lm} j_{l}(\sqrt{\lambda_l} R)}{(2l+1)R}
+\frac{\lambda_l{\tilde c}^{r}_{lm}\left(rj_{l}(\sqrt{\lambda_l} r)\right)'|_{r=R} }{(2l+1)lR}=
- R^{l-1} .
\end{eqnarray}
Since $p_o$ is harmonic then for matching $p=r^lY_{lm}(\theta, \phi)$ on $S$ it has the form $p_o=R^{2l+1}Y_{lm}(\theta, \phi)/r^{l+1}$.
Consequently, the boundary condition in Eq.~(\ref{bs}) results after rearrangements and using Eq.~(\ref{sd}) in the following equation (since ${\tilde c}^{r}_{lm}$ do not depend on the index $m$ then we omit it):
\begin{eqnarray}&&\!\!\!\!\!\!\!\!\!\!\!\!\!\!
{\tilde c}^{r}_{l}=\frac{(l+1)R^l}{\lambda_l j_{l}(\sqrt{\lambda_l} R)}. \label{coeff}
\end{eqnarray}
Summing the last two equations we find that $x_l\equiv \sqrt{\lambda_l} R$ satisfies
\begin{eqnarray}&&\!\!\!\!\!\!\!\!\!\!\!\!\!\!\!
\left[\ln(x_lj_{l}(x_l))\right]'
=-\frac{(3l+2)l}{(l+1)x_l}.
\end{eqnarray}
We find using Eq.~(\ref{or}) that the smallest eigenvalue $x_1$ obeys
\begin{eqnarray}&&\!\!\!\!\!\!\!\!\!\!\!\!\!\!\!
2x_1^2\sin x_1=3(x_1\cos x_1-\sin x_1),\ \  x_1\approx 3.506.
\end{eqnarray}
Since $x_1$ is smaller than the smallest eigenvalue associated with the other see of eigenmodes, see Eq.~(\ref{fx}), then it provides for the smallest eigenvalue giving that for a ball of radius $R$ we have
\begin{eqnarray}&&\!\!\!\!\!\!\!\!\!\!\!\!
\frac{\int_V \left(\nabla \times \bm B\right)^2d\bm x}{\int_V  B^2 d\bm x}\geq \frac{x_1^2}{R^2},\ \  x_1\approx 3.506.  \label{re1}
\end{eqnarray}
The RHS is larger than that in Eq.~(\ref{re}) since the denominator in the latter is larger than the one in Eq.~(\ref{re1}). This inequality has never been obtained previously.

\subsection{Structure of minimal dissipation modes}\label{sec:fe}

\begin{figure}
 \centerline{\includegraphics[scale=0.3]{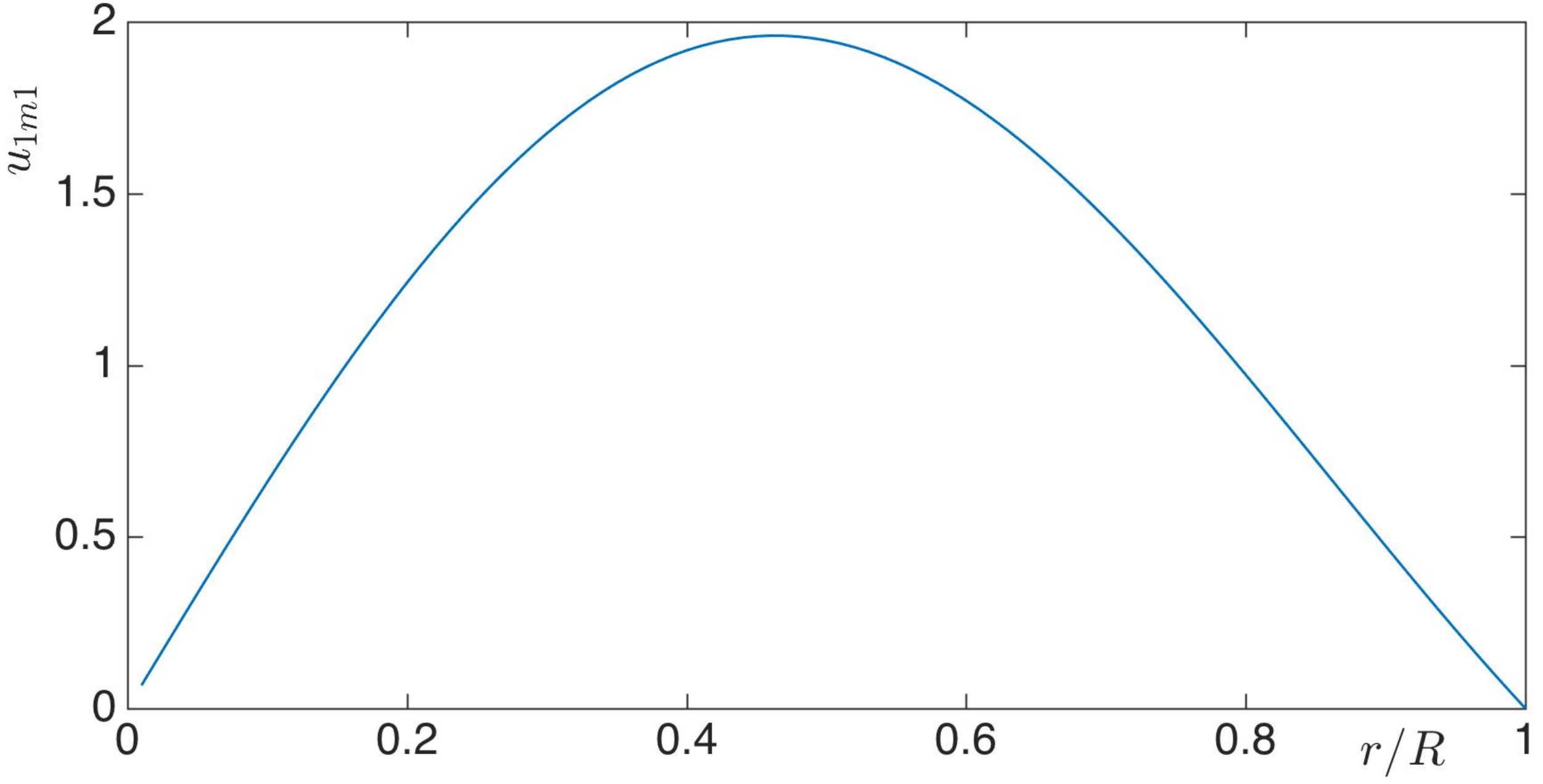}}
 \caption{The radial structure of the slowest decaying mode of unsteady Stokes flow in a ball.}
\label{fig:u}
\end{figure}

\begin{figure}
 \centerline{\includegraphics[scale=0.3]{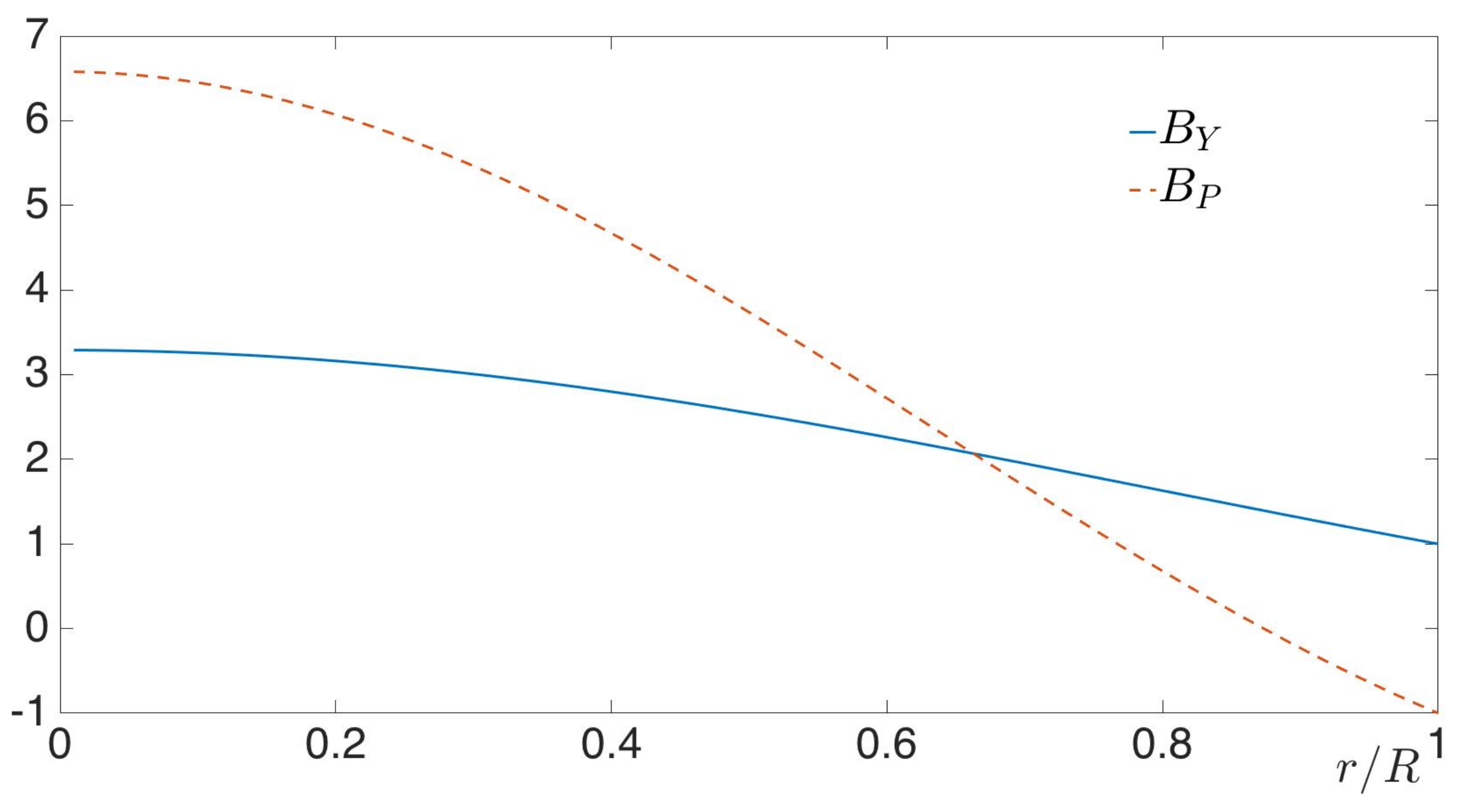}}
 \caption{The radial structure of the components $B_{Y}$ and $B_{P}$ of the slowest decaying mode of the induction equation in the absence of
flow in a ball ($B_Y$ and $B_P$ are the coefficients of $\bm Y_{1m}$ and $\bm \Psi_{1m}$, respectively).}
\label{fig:b}
\end{figure}

Before moving on to further expanding the stability investigations it is of interest to briefly dwell on the structure of the free decay modes found in the previous subsections. Thus, three different solenoidal modes that minimize dissipation in a ball of radius $R$ have been observed. One mode is the flow $\bm u$ that minimizes $\int_V \left(\nabla \times\bm u\right)^2d\bm x/\int_V u^2 d\bm x$ and
vanishes on $S$. The other two modes are admissible magnetic fields, in the sense described in Sec. \ref{adm}, that minimize $\int_V \left(\nabla \times\bm B\right)^2d\bm x/\int B^2 d\bm x$ and
$\int_V \left(\nabla \times\bm B\right)^2d\bm x/\int_V B^2 d\bm x$ respectively. The first two modes are given
by (we use Eq.~(\ref{or}) and change overall multiplicative constants)
\begin{eqnarray}&&\!\!\!\!\!\!\!\!\!
\bm u_{1m1}=\left(\frac{R}{xr}\sin\left(\frac{x r}{R}\right)-\cos\left(\frac{x r}{R}\right)\right)\frac{xR\bm \Phi_{1m}}{r},\label{mois}\\&&\!\!\!\!\!\!\!\!\!
x=\tan x\approx 4.493;\ \ \bm B_{1m1}\!=\!\left(\frac{\sin y}{y}\!-\!\cos y\right)\frac{\bm Y_{1m}}{y^2}
\nonumber\\&&\!\!\!\!\!\!\!\!\!
\!+\!\left(\cos y\!+\!\left(y\!-\!\frac{1}{y}\right)\sin y\right)\frac{\bm \Psi_{1m}}{2y^2},\ \ y\equiv \frac{\pi r}{R},
\end{eqnarray}
where $\bm u_{1m1}(r=R)=0$. These modes have zero pressure, defined as the Lagrange multiplier associated with the solenoidality condition. Thus the variations of
$\int_V \left(\nabla \times\bm u\right)^2d\bm x/\int_V u^2 d\bm x$ and $\int_V \left(\nabla \times\bm B\right)^2d\bm x/\int_V B^2 d\bm x$, considered over $\bm u$ and $\bm B$ unconstrained by the solenoidality,
vanish for $\bm u=\bm u_{1m1}$ and $\bm B=\bm B_{1m1}$, respectively. We have
\begin{eqnarray}&&\!\!\!\!\!\!
\frac{\int_V \left(\nabla \times\bm u_{1m1}\right)^2d\bm x}{\int_V u_{1m1}^2 d\bm x}=\frac{x^2}{R^2},\ \  \frac{\int_V \left(\nabla \times \bm B_{1m1}\right)^2d\bm x}{\int  B_{1m1}^2 d\bm x}= \frac{\pi^2}{R^2}.\nonumber
\end{eqnarray}
There is degeneracy in $m$ so that the equalities hold for $m=\pm 1$ or $m=0$.
The modes have direct physical meaning. The flow $\bm u_{1m1}$ is the slowest decaying mode of unsteady Stokes flow in a ball and the magnetic field $\bm B_{1m1}$ is the slowest decaying mode of the induction equation in the absence of
the flow, see above and \cite{backus}. The mode $\bm u_{1m1}$ vanishes at $r=0$ and $r=R$. In contrast, $\bm B_{1m1}$ is non-zero at these $r$. We plot $u_{1m1}(r)$ and $B_{1m1}(r)$ in Figs. \ref{fig:u} and \ref{fig:b}, respectively.
It is seen that $\bm B_{1m1}$ is appreciable on the boundary of the fluid domain so that this mode uses that the magnetic field spreads beyond the fluid domain for lowering the dissipation.

It might be thought that zero pressure is an inherent property of this type of variational problems, cf. \cite{backus}. However the magnetic field $\bm B_0$ that minimizes
$\int_V \left(\nabla \times\bm B\right)^2d\bm x/\int_V B^2 d\bm x$ has a non-trivial pressure given by $rY_{1m}(\theta, \phi)$ (pressure is a linear function of coordinates). Thus it minimizes a linear
combination of $\int_V \left(\nabla \times\bm B\right)^2d\bm x/\int_V B^2 d\bm x$ and $\int rY_{1m} \nabla\cdot\bm Bd\bm x$ over $\bm B$ unconstrained by the solenoidality (and not simply
$\int_V \left(\nabla \times\bm B\right)^2d\bm x/\int_V B^2 d\bm x$). We have
\begin{eqnarray}&&\!\!\!\!\!\!\!\!\!
\bm B_0\!=\!\left(\frac{2R j_{1}(x_1 r/R)}{r j_{1}(x_1)}
\!+\!1\right)\bm Y_{1m}
\!+\!\left( \frac{R\left(rj_{1}(x_1 r/R)\right)' }{rj_{1}(x_1)}
\!+\!1\right)\nonumber\\&&\!\!\!\!\!\!\!\!\!  \times \bm \Psi_{lm};\ \ \frac{\int_V\!\! \left(\nabla\! \times\! \bm B_{1m1}\right)^2d\bm x}{\int \!\! B_{1m1}^2 d\bm x}\!= \!\frac{x_1^2}{R^2},\ \ x_1\approx 3.506, \nonumber
\end{eqnarray}
where we used Eq.~(\ref{coeff}) and changed the overall multiplicative constant. This formula can be rewritten via the trigonometric functions, similarly to Eqs.~(\ref{mois}), by using Eq.~(\ref{or}). The
field $\bm B_0$ is appreciable on the boundary, similarly to $\bm B_{1m1}$.

The last sections prove Eqs.~(\ref{ins}) finishing the derivation of the non-linear stability criteria given by Eqs.~(\ref{crt}). These are given via the properties of the strain and not $\bm v_0$ itself.
In the next section we consider derivation of stability criteria in terms of the velocity.

\section{Other boundary conditions}\label{obc}

The criteria that are displayed in Eqs.~ (\ref{crt}) and (\ref{criteria}) have been derived under the assumption of ``true" physical boundary conditions, namely the normal component of the velocity is zero
(thus generalising results of \cite{backus} that have been obtained for rigid rotation at most), as well as continuity of the magnetic field at the boundary, and the requirement that $r^3{\bm B}$ is finite in all space.
Significant simplification of the calculations may result from the assumption that $T'({\bm B})=T({\bm B})$ in Eq.~(\ref{landau}), namely that $({\bm v}_0\cdot{\bm B})({\hat {\bm n}}\cdot{\bm B})=0$ on the boundary.
This rather fits with boundary conditions that are frequently used for astrophysical numerical simulations where either the normal component of the magnetic field or its tangential component are assumed to vanish on the
computational domain boundary. Furthermore, numerical simulations indicate that the generated magnetic field in a sphere concentrates near the origin and does not extend all the way to the boundaries \citep{herreman,chen}, cf.
however the structure of the minimal Joule heating modes above. Finally some phenomena, such as kinematic dynamo in homogeneous turbulence \cite{xl}, happen in the bulk and do not depend on the boundaries. In this section we
explore the effect of using other boundary conditions by employing the assumption that all boundary terms that occur in integrations by parts can be neglected as modelled by $\bm B(S)=0$. We will see that the calculations
simplify most significantly in comparison with the previous sections.

The assumption that $T'({\bm B})=T({\bm B})$ implies that $Q_1=Q_2$ in Eq.~(\ref{landau}) up to the change of sign of $s$ in the definition of $T$. This allows to simply transfer the results obtained for $Q_1$ to $Q_2$
by changing $\bm v_0$ by $-\bm v_0$. The criterion of global stability given by Eq.~(\ref{crt}) remains intact under this change. The criterion given by Eq.~(\ref{criteria}) becomes
\begin{eqnarray}&&\!\!\!\!\!\!\!\!\!\!\!\!\!\!
\frac{v_{0max}R}{\nu}<x,\ \ \frac{v_{0max}R}{\eta}<x,\ \ x\approx 4.493.\label{soot}
\end{eqnarray}
According to the above criteria global stability is guaranteed at smaller $\eta$ than in the case of the true boundary conditions. This is so since the condition $\bm B(S)=0$ cuts off modes that grow under true boundary conditions,  a fact to be reckoned with in practice.

It is sometimes possible to derive more detailed criteria. We consider as an example the Couette flow that possesses the symmetry $\bm v_0\to -\bm v_0$ (more generally this symmetry holds for flows for which the non-linear term
in the Navier-Stokes equations vanishes identically such as pipe or channel flows). This flow holds between two concentric cylinders with inner and outer radii $R_1$ and $R_2$ and angular velocities
$\Omega_1$ and $\Omega_2$, respectively. We find immediately by applying the results of \cite{serrin} that if both $\nu$ and $\eta$ are greater than $\nu_c$ defined by
$\nu_c^{-1}|\Omega_2-\Omega_1|=\pi^2(R_2^2-R_1^2)\left(R_1R_2\ln(R_2/R_1)\right)^{-2}$
then arbitrary disturbances decay. Thus in the limit of large viscosity and diffusivity (here given by $\nu>\nu_c$ and $\eta>\nu_c$) the laminar hydrodynamic Couette flow is stable.

Further transfer of known results for $Q_1$ to $Q_2$ is straightforward. For instance using Eq.~(\ref{inq1}) it is found that
\begin{eqnarray}&&\!\!\!\!\!\!\!\!\!\!\!\!\!\!
\frac{\dot E}{2}\!\leq\! \!\left(\left|min_V[eig\left(s\right)] \right|\!-\!\frac{\alpha \nu}{d^2}\right)E_{kin}
\nonumber\\&&\!\!\!\!\!\!\!\!\!\!\!\!\!\!
\!+\!\left(max_V[eig\left(s\right)]\!-\!\frac{\alpha \eta}{d^2}\right)E_m
\nonumber\\&&\!\!\!\!\!\!\!\!\!\!\!\!\!\!
\leq  \ max \left(\left|min_V[eig\left(s\right)] \right|\!-\!\frac{\alpha \nu}{d^2}, max_V[eig\left(s\right)]\!-\!\frac{\alpha \eta}{d^2}\right)E.\nonumber
\end{eqnarray}
We conclude that the logarithmic decay rate of the energy is no smaller than $2 \left| max \left(\left|min_V[eig\left(s\right)] \right|\!-\!\alpha \nu/d^2, max_V[eig\left(s\right)]\!-\!\alpha \eta/d^2\right)\right|$.
This reduces for $\bm B\equiv 0$ to the result of \cite{serrin}. In our approach the above inequalities would include $x^2/R^2$ instead of $\alpha/d^2$, see Eqs.~(\ref{crt}). This would be a stronger result assuming the
validity of the generalized Rayleigh conjecture. Similarly usage of result of \cite{serrin} for $Q_1$
proves that the flow $\bm v_0$ is globally stable if
\begin{eqnarray}&&\!\!\!\!\!\!\!\!\!\!\!\!\!\!
\frac{v_{0max}d}{\nu}< \sqrt{\alpha},\ \ \frac{v_{0max}d}{\eta}< \sqrt{\alpha},\ \ \sqrt{\alpha}\approx 5.71.\label{root}
\end{eqnarray}
This result is weaker than Eq.~(\ref{soot}) if the generalized Rayleigh conjecture holds. If the conjecture is not assumed then $R$ is the smallest radius of a ball enclosing the domain and the result resembles the linear
Childress and Backus bounds considered later.
%
%
%
%
%

\section{Non-linear decay of magnetic field at high Reynolds number} \label{tr}

If the kinetic Reynolds number is high, perturbations about $\bm v_0$ (concentrated at scales larger than dissipation scales) will grow, including those caused by a seed magnetic field. Therefore the distance between the MHD solutions $\bm v$, $\bm B$ and $\bm v_0$, as measured by $E$, would grow and $E$ is not a good measure for determining whether the magnetic perturbations disappear. Instead we can consider if the magnetic component $E_m$ of $E$ decays, guaranteeing that any initial disturbance $\bm v(t=0)$, $\bm B(t=0)$ will decay under the non-linear MHD evolution to a purely hydrodynamic flow with
$\bm B=0$. In this section we derive the criterion for this decay. We use the physical boundary conditions with $\bm B$ extending over the entire space.

Our previous considerations apply mostly to the low Reynolds number laminar stationary background flows though generalizations are possible for stable time-dependent flows. In the opposite limit of large Reynolds number,
when the flow is not hydrodynamically stable and turbulent \citep{frisch,ll6}, the Reynolds-Orr equation and its magnetic extension become less relevant because $Q_1(\bm u)$ in Eq.~(\ref{landau}) is positive for a vast range of
$\bm u$. Still the magnetic Reynolds number might be so small that time-derivative of magnetic energy, given by Eq.~(\ref{magnetoc}), is negative for all $\bm B$ for reasonable properties of $\bm v_0$
(large $Re$ and small $Re_m$ implies small magnetic Prandtl number $\nu/\eta$).

We observe that time derivative of the magnetic energy $E_m$, given by Eq.~(\ref{start}), equals the derivative of $E$ in Eq.~(\ref{fedor}) with terms involving $\bm u$ omitted and $\bm v_0$ replaced by $\bm v$.
We find the inequality
\begin{eqnarray}&&\!\!\!\!\!\!\!\!\!\!\!\!\!\!
\frac{dE_m}{dt}
\leq \int_V\!\! \left(\frac{(\bm B\times \bm v)^2}{2{\tilde \kappa}}+\frac{{\tilde \kappa}j^2}{2}\!-\!\eta j^2\right)d\bm x ,
\end{eqnarray}
which is obtained from Eq.~(\ref{kappa}) by omitting the contribution of $\bm u$ and using $\bm v$, that depends on $\bm B$ via the non-linear MHD equations given by Eqs.~(\ref{magnc}), instead of $\bm v_0$. We find using
${\tilde \kappa}=\eta$ (we saw that this is not the optimal choice of ${\tilde \kappa}$ however it simplifies calculations. The resulting bound is not the tightest.) that
\begin{eqnarray}&&\!\!\!\!\!\!\!\!\!\!\!\!\!\!
\frac{d\ln E_m}{dt}=\frac{2}{\int B^2d\bm x}\frac{dE_m}{dt}
\leq  \frac{v^2_{max}\int_V B^2d\bm x}{\eta \int B^2d\bm x}-\frac{\eta\int j^2 d\bm x}{\int B^2d\bm x}
\nonumber\\&&\!\!\!\!\!\!\!\!\!\!\!\!\!\!
\leq  \frac{v^2_{max}}{\eta}-\frac{\eta \pi ^2}{R^2},\label{sas}
\end{eqnarray}
where we used Eq.~(\ref{re}) and $v_{max}$ is the maximal value of $\bm v$ over the fluid domain.
This equation holds instantaneously at time $t$ where for turbulence $v^2_{max}(t)$ is random yet bounded. Thus, despite that $\bm v$ depends in an unknown fashion on $\bm B$, the equation implies that for any given
disturbance the magnetic energy decays instantaneously for $\eta$ above a threshold whose value depends on the disturbance's details. Information on the global behavior can be obtained by integrating the above equation
over time, dividing by $t$, and taking $t\to\infty$,
\begin{equation}
\lim_{t\to\infty}\frac{\ln (E_m(t)/E_m(0))}{t}\!\leq\!\frac{\left\langle v^2_{max}\right\rangle}{\eta}\!-\!\frac{\eta \pi^2}{R^2}, \label{log}
\end{equation}
where the angular brackets stand for time averaging.

The above equation allows to establish the non-linear stability bound by consistency demand.
If we assume that the flow is unstable with respect to generation of magnetic field then the asymptotic state at large times is a stationary MHD turbulence where we assume that the driving forces are stationary.
Therefore $E_m(t)$ is finite and Eq.~(\ref{log}) gives
\begin{equation}
\frac{R\sqrt{\left\langle v^2_{max}\right\rangle} }{\eta \pi}\!\geq \!1. \label{cf}
\end{equation}
We conclude that $E_m(t)$ cannot remain finite and must decay unless the Reynolds number defined by the LHS of the above equation is larger than one. That Reynolds number involves a property of the
stationary MHD flow, $\left\langle v^2_{max}\right\rangle$. Its value is unknown however it is equal by order of magnitude to the kinetic energy $\left\langle v^2\right\rangle/2$. Typically this would also agree
by order of magnitude with the purely
hydrodynamic turbulence's  $\left\langle v_0^2\right\rangle/2$, determined by Eq.~(\ref{nc}), or the square of the integral scale velocity of turbulence $U^2$, see e.g. \cite{frisch}. Hence if we introduce a dimensionless coefficient $c$ by $\left\langle v^2_{max}\right\rangle=c^2U^2$ then we find the global stability criterion
\begin{eqnarray}&&\!\!\!\!\!\!\!\!\!\!\!\!\!\!
Re_m\!<\!\frac{\pi}{c},\ \ Re_m\equiv \frac{UR}{ \eta}, \label{sde}
\end{eqnarray}
where $Re_m$ is a possible definition of the usual magnetic Reynolds number. The only difference of this criterion from the last of Eqs.~(\ref{root}), besides the obvious change of the definition of $Re_m$, is the presence of unknown constant $c$ which is plausibly of order one. If we want to avoid the
presence of unknown factors then we must either perform accurate study of $\left\langle v^2_{max}\right\rangle$ or limit the consideration to small initial perturbations of the magnetic field which constitutes the kinematic dynamo
problem. The point of introducing $c$ is that the criterion is given in terms of hydrodynamic turbulence and not MHD turbulence.

Some comments are in order. The bound seemingly could be tightened by using
\begin{eqnarray}&&\!\!\!\!\!\!\!\!\!\!\!\!\!\!
\frac{2}{\int_V B^2d\bm x}\frac{dE_m}{dt}
\leq  \frac{v^2_{max}}{\eta}-\frac{\eta\int j^2 d\bm x}{\int_V B^2d\bm x}\leq  \frac{v^2_{max}}{\eta}-\frac{\eta x_1^2}{R^2},
\end{eqnarray}
where $x_1\approx 3.506$, instead of Eq.~(\ref{sas}). This would give factor of $x_1$ instead of $\pi$ in Eq.~(\ref{sde}). We will not explore this bound that demands more cumbersome calculations. A stability condition similar to Eq.~(\ref{sde}) could be derived in terms of the eigenvalues of the strain matrix, cf. Eq.~(\ref{crt}).
Proceeding as in previous sections we find starting from Eq.~(\ref{un}) that
\begin{eqnarray}&&\!\!\!\!\!\!\!\!\!
\frac{d\ln E_m}{dt}\!\leq \!2\ max_V[eig\left({\tilde s}\right)]\!-\!\frac{2\eta\int_V (\nabla\!\times\!\bm B)^2d\bm x}{\int \!\! B^2 d\bm x}
\nonumber\\&&\!\!\!\!\!\!\!\!\!
\leq
2\ max_V[eig\left({\tilde s}\right)]\!-\!\frac{2\eta \pi^2}{R^2},
\end{eqnarray}
where ${\tilde s}$ is the strain matrix of $\bm v$ i.e. ${\tilde s}_{ik}=(\nabla_iv_k+\nabla_kv_i)/2$. We find as previously that self-consistency of the assumption that magnetic energy does not decay demands
\begin{equation}
\frac{R^2\left\langle max_V[eig\left({\tilde s}\right)]\right\rangle }{\eta \pi^2}\!\geq \!1,
\end{equation}
cf. Eq.~(\ref{cf}). However, since turbulent velocity gradients are determined by small-scale eddies then this bound is weaker than
Eq.~(\ref{sde}) by a power of $Re$ that is $R\left\langle max_V[eig\left({\tilde s}\right)]\right\rangle/\sqrt{\left\langle v^2_{max}\right\rangle}$ is a power of $Re$. Yet the criterion can be of use for flows with moderate
Reynolds numbers. Finally, non-linear stability criteria can also be derived for other boundary conditions, cf. the previous section.

\section{Representation of the energy growth rate for the kinematic dynamo problem} \label{kinemtic}

The criteria of the previous section can be given in terms of the properties of hydrodynamic turbulence without any unknown factors in the framework of the kinematic dynamo.
The evolution of the magnetic field $\bm B$ is then determined by:
\begin{eqnarray}&&\!\!\!\!\!\!\!\!\!\!\!\!\!\!
\partial_t\bm B+(\bm v_0\cdot \nabla) \bm B=(\bm B\cdot \nabla) \bm v_0+\eta\nabla^2 \bm B. \label{magn}
\end{eqnarray}
This equation is the second of Eqs.~(\ref{magnc}), under the assumption that the Lorentz force $(\bm B\cdot\nabla)\bm B$ term in the momentum equation is negligible due to the smallness of $\bm B$, so that $\bm v$
becomes $\bm v_0$ given by Eq.~(\ref{nc}). Thus $\bm v_0$ in Eq.~(\ref{magn}) is a given flow that is independent of $\bm B$. We are interested in the asymptotic properties of the evolution at large times. It is seen immediately
that considerations of the previous section apply to $E_m$ in this case with $\bm v$ replaced by $\bm v_0$. This implies that if any of the conditions
\begin{equation}
\frac{R\sqrt{\left\langle v^2_{0max}\right\rangle} }{\eta \pi}\!< \!1,\ \ \frac{R^2\left\langle max_V[eig\left(s\right)]\right\rangle }{\eta \pi}\!< \!1,\label{crp}
\end{equation}
the magnetic energy decays. These are known criteria of linear stability, the Childress \citep{leeds} and Backus \cite{backus} bounds, respectively. These are given here in the form suited for a time-dependent flow including
turbulence. The bounds apply also to laminar flows where the averaging can be omitted. If our generalized Rayleigh conjecture holds then the Childress bound can be tightened. Our considerations also demonstrate that the assumption
made in the derivation of \cite{backus} that the flow is a rigid rotation on $S$ can be relaxed.

It is seen from the previous section that logarithmic growth rate is a quantity which has simple properties. We study the rate using the boundary condition $\bm B(S)=0$ so that the magnetic field is confined to the volume of
the fluid. We use the energy density $e_m\!\equiv \!\int_V B^2d\bm x/(2V)$, rather than the energy itself.
We observe that then Eq.~(\ref{magnetoc}) yields:
\begin{eqnarray}&&\!\!\!\!\!\!\!\!\!\!\!\!\!\!
\lambda(t) \equiv  \frac{1}{2}\frac{d \ln e_m}{dt}\!=\!\frac{1}{V}\int\!\! \left(\bm b s\bm b\! -\!\eta \left(\nabla \bm b\right)^2 \right) d\bm x=\frac{Q_2(\bm b)}{V},
\label{ios1}
\end{eqnarray}
where $\lambda(t)$ is a logarithmic growth rate of the magnetic field and we introduced the "spatially normalized magnetic field" $\bm b(t, \bm x)\equiv \bm B(t, \bm x)/\sqrt{2e_m(t)}$. It is readily seen that $\bm b$ obeys
$\int b^2 d\bm x/V \!=\!1$ and
\begin{eqnarray}&&\!\!\!\!\!\!\!\!\!\!\!\!\!\!
\partial_t\bm b\!+\!(\bm v_0\!\cdot\! \nabla) \bm b\!=\!(\bm b\!\cdot\! \nabla) \bm v_0
\!-\!\bm b\nabla\!\cdot\! \bm v_0\!+\!\eta\nabla^2 \bm b\!-\!\lambda(t)\bm b,\label{ios}
\end{eqnarray}
where $\nabla\cdot \bm b=0$ and we used Eq.~(\ref{magn}). Previous considerations for $Q_2$ result in the following inequalities
\begin{eqnarray}&&\!\!\!\!\!\!\!\!\!\!\!\!\!\!
\lambda \leq\! max_V[eig\left(s\right)]\!-\!\frac{\eta x^2}{R^2},\ \ \lambda \leq\! \frac{v^2_{0max}}{2\eta}\!-\!\frac{\eta x^2}{2R^2}, \label{inequalities}
\end{eqnarray}
which holds instantaneously at time $t$ and differ from the equations of the previous section by factor of $x$ instead of $\pi$ due to different boundary conditions on $\bm B$. We also find another inequality by introducing dimensionless variables as previously and using
"time-dependent magnetic Reynolds number" (the time-dependence comes from $T(\bm b)$,
\begin{eqnarray}&&\!\!\!\!\!\!\!\!\!\!\!\!\!\!
\lambda(t)
\leq\! - \frac{D(\bm b)}{V Re_m}\left(1-\frac{Re_m}{{\tilde Re}_m(t)}\right),\nonumber\\&&\!\!\!\!\!\!\!\!\!\!\!\!\!\!
\frac{1}{{\tilde Re}_m(t)}\equiv max_{\bm b}\left(-\frac{T(\bm b)}{D(\bm b)}\right), \label{inequali}
\end{eqnarray}
where the forms $T$ and $D$ are defined in Eq.~(\ref{landau}).

\subsection{Laminar flow}

A simple use of the above equations is provided by the kinematic dynamo problem for a time-independent laminar flow $\bm v_0$. In this case Eq.~(\ref{magn}) implies that at large times the magnetic field is given by
$\bm B\propto \exp(\lambda t)\bm b(\bm x)$ where $\lambda$ is the largest eigenvalue of the time-independent evolution operator in Eq.~(\ref{magn}). Then Eq.~(\ref{ios1}) provides the connection between $\lambda$ and $\bm b$.

We also observe that in the case of kinematic dynamo problem for laminar flow all the results that were derived previously from the extended Reynolds-Orr equation hold with $Q_1$ omitted. Some conclusions are obtainable from observing that $Q_1$ and $Q_2$ interchange under the change of sign of $\bm v_0$. Thus if a hydrodynamic study proves that $Q_1$ is negative definite for a certain range of
$Re$, and the flow has $\bm v_0\to-\bm v_0$ symmetry, then $Q_2$ is negative definite for the same range of $Re_m$ (with corresponding implications for stability).

We have obvious conclusions from the inequalities given by Eqs.~(\ref{inequalities}), (\ref{inequali}). Since for the used boundary condition $\bm B(S)$ the forms $Q_i$ agree, up to sign-reversal of $\bm v_0$, then we
could similarly use the inequalities of \cite{serrin} derived for $Q_1$. For instance the last of Eqs.~(\ref{inequalities}) gives that the magnetic field decays if $v_{0max}R/\eta<x$ whereas the usage of \cite{serrin}
would give as a criterion $v_{0max}d/\eta<\sqrt{\alpha}\approx 5.71$.
Similarly, Eq.~(\ref{inequali}) gives the stability bound $Re_m<{\tilde Re}_m$.

\subsection{Turbulence.} We return to the general case of time-dependent turbulent flow. We have from Eq.~(\ref{ios1})
\begin{eqnarray}&&\!\!\!\!\!\!\!\!\!\!\!\!\!\!\!
\lim_{t\to\infty}\!\frac{1}{2t}\ln \left(\frac{e_m(t)}{e_m(0)}\right)\!=\!\langle \lambda\rangle\!=\!\int\!\! \left(\!\langle\bm b s \bm b\rangle\! -\!\eta \left\langle\left(\nabla \bm b\right)^2\right\rangle
\!\right)\!\frac{d\bm x}{V},\label{land}
\end{eqnarray}
provided that the limit exists where we remind the reader that the angular brackets stand for time-average. We find from Eq.~(\ref{inequali}) that if $Re_m<{\tilde Re}_m(t)$ at all times then the energy decays.
Universal stability conditions are found readily from Eqs.~(\ref{inequalities}) that give
\begin{eqnarray}&&\!\!\!\!\!\!\!\!\!\!\!\!\!\!
\langle \lambda\rangle\leq\! \langle max_V[eig\left(s\right)]\rangle \!-\!\frac{\eta x^2}{R^2},\ \ \langle \lambda\rangle \leq\! \frac{\langle v^2_{0max}\rangle}{2\eta}\!-\!\frac{\eta x^2}{2R^2}.
\end{eqnarray}
This gives simple anti-dynamo theorems, similar to those in Eqs.~(\ref{crp}) with the difference due to the boundary conditions. If we define the strain-based Reynolds numbers that are based on the gradients and the velocity as $Res_m\equiv
\langle max_V[eig\left(s\right)]\rangle R^2/\eta$ and $Re_m\equiv \langle v^2_{0max}\rangle^{1/2} R/\eta$ then if either of the conditions
\begin{eqnarray}&&\!\!\!\!\!\!\!\!\!\!\!\!\!\!
Res_m<x^2\approx 20.19,\ \ Re_m<x\approx 4.493,\label{fd}
\end{eqnarray}
holds, small perturbations of the magnetic field decay. For single-scale homogeneous random flow the first of the above criteria is similar to that obtained from theoretical studies that are based on the
Kazantzev-Kraichnan model of turbulence \citep{xl}. For turbulence with fluctuations at many scales, the flow gradients are determined by small-scale eddies \citep{frisch}. Thus $\langle max_V[eig\left(s\right)]\rangle$ is larger
than $\langle v^2_{0max}\rangle^{1/2}/R$ by a power of Reynolds number and only the last of Eqs.~(\ref{fd}) is really useful. We observe that $Re_m$ is similar to the usual way of defining the magnetic Reynolds number
as the product of the integral scale $L$ and the integral scale velocity $U$ divided by $\eta$. Indeed, in many practically relevant cases turbulence is stirred on the scale of the container so that $L\sim R$. Moreover
the statistics at the integral scale is not intermittent so that $\langle v^2_{0max}\rangle^{1/2}$ would usually be of the same order as $U$ (in many cases both velocities are similar to the velocity difference prescribed
at the boundary). We conclude that $Re_m$ is of order of the usual magnetic Reynolds number which allows comparison with the usual criteria \citep{xl}.

Finally we provide a more detailed criterion by starting from the assumption that at large times the magnetic energy spectrum $E_B(t, k)$ of the magnetic field is separable, $E_B(t, k)=e_m(t)e_B(k)$, where $k$ is the wavenumber.
In the case of dynamo, where the field grows, this form is observed in a number of simulations of homogeneous isotropic turbulence, see \cite{xl} and references therein. Since the magnetic field energy is $\int E_B(t, k) dk$ then
we have $\int e_B(k) dk=1$. The spectrum of $\bm b$ is seen from the definition in Eq.~(\ref{ios}) to be given by $e_B(k)/2$. We find
\begin{eqnarray}&&\!\!\!\!\!\!\!\!\!\!\!\!\!
\langle \lambda\rangle\!\leq\!\! (2\eta)^{-1}\left(\left\langle v_{0max}^2\right\rangle\!-\!\eta^2\left\langle\! \left(\nabla \bm b\right)^2\!\right\rangle_c\right)
\nonumber\\&&\!\!\!\!\!\!\!\!\!\!\!\!\!
=\eta\left(Re_m^2-1\right)\int_0^{\infty}\!\!\!\! k^2 e_B(k)dk/4,
\end{eqnarray}
where we designate by angular brackets with subscript $c$ the spatio-temporal averages that can be assumed to be equal to the ensemble averages by ergodicity. We introduced the magnetic Reynolds number
$Re_m^2\equiv 2\left\langle v_{0max}^2\right\rangle/\left(\eta^2 \int_0^{\infty} k^2 e_B(k)dk\right)$. The difference of the inequality above from those used previously is that we treat the energy dissipation term without
approximations. We find that for $Re_m<1$ we have $\langle \lambda\rangle<0$ and there is no dynamo effect. We observe that if we introduce a correlation length $l_c$ of the magnetic field by
$2l_c^{-2}\equiv \int_0^{\infty} k^2 E_B(t, k)dk/\int_0^{\infty}  E_B(t, k)dk=\int_0^{\infty} k^2 e_B(k)dk$ then $Re_m=l_{c}\sqrt{\left\langle v_{0max}^2\right\rangle}/\eta$.
Since near the transition the correlation length of the magnetic field is of order of $L$ then the above definition agrees by order of magnitude with the more usual definition $Re_m=U L/\eta$. This implies that for $UL/\eta\ll 1$ there is no dynamo.

\section{Conclusions}
Our work describes what can be inferred about the magnetic field generation in conducting fluids from general principles, without reverting to detailed calculations, theoretical or numerical. This includes thresholds for
non-linear instability, that seemingly were not considered in this context previously. This is done by providing an extension of the Reynolds-Orr equation to magnetohydrodynamics. The latter describes the evolution of the total
energy, both kinetic as well as magnetic and hence, unlike previous works, enables the derivation of lower bounds for dynamo action for both the Reynolds number as well as for the magnetic Reynolds number.  The extended Reynolds-Orr
equation has been derived for boundary conditions of zero normal velocity (and arbitrary tangential velocity) and the continuous connection of the magnetic field to a decaying external potential field. We demonstrated that
the extension has similar powerful implications for the non-linear stability as its hydrodynamic counterpart. Thus we can prove that there is a range of small hydrodynamic and magnetic Reynolds numbers for
which the laminar flow without magnetic field is stable with respect to disturbances of arbitrary magnitude.

We have started by deriving lower bounds for the logarithmic decay rates of kinetic energy of the fluid and the Joule heating of non-flowing conducting medium in a given domain. We introduced conjecture that the domain shape that
creates optimal conditions for minimization of the dissipative processes has maximal symmetry i.e. a sphere. This generalizes from scalar to solenoidal vector fields the proved conjecture by Rayleigh and leads to a generalised
Rayleigh-Faber-Krahn inequality. The extremal modes of dissipation were constructed explicitly for a ball by using vector spherical harmonics. An interesting side result emerges from these calculations, that the slowest
decaying mode has zero pressure. Whether this is true for a domain of any shape is left for future work. We used these results in order to derive lower bounds for the stability of hydrodynamic as well as magnetic perturbations
(not necessarily small). The lower bounds are expressed in terms of the Reynolds as well as magnetic Reynolds numbers that are based on the maximal velocity of the unperturbed flow or eigenvalues of strain.
We have also provided proved weaker bounds that do not involve any conjecture. Finally, we considered also other boundary conditions, that occur in numerical work, demonstrating that they lead to significantly simpler calculations and qualitatively similar bounds.

In the case of turbulence we provided a criterion for the decay of arbitrarily large initial fluctuations of the magnetic field. This is given via properties of hydrodynamic turbulence and an unknown coefficient that is
probably of order one. Here more detailed studies are needed including that, to the best of our knowledge, the critical magnetic Reynolds number below which initial, possibly large, disturbances of the magnetic field decay,
is yet to be obtained numerically (similar studies for small initial disturbances are well-known \cite{xl}). Our result does prove that there is a finite threshold value of the magnetic Reynolds number below which the magnetic
field decays.

We remark that the derived stability criteria, are probably significantly smaller than the actual sharp bounds that would be found from detailed studies that are usually numerical \citep{xl, herreman, chen}. This
parallels the situation for the estimates of the critical Reynolds number of hydrodynamic instability using the Reynolds-Orr equation - the estimates are too conservative \citep{ll8,serrin}. However the described
approach seems to be the only theoretical route to non-linear stability which is valuable since linear stability analysis is intrinsically incomplete. Moreover, even in the case of linear stability, the provided estimates are valuable
since they do not require extensive calculations for reaching conclusions of wide generality. In this context it is worth stressing that the considerations above do not assume homogeneous isotropic turbulence and could be used
for other random flows including the less studied inhomogeneous turbulence.

We also demonstrated that the logarithmic growth rate of magnetic field in the kinematic dynamo problem is described by an equation similar to that in the Reynolds-Orr equation.
We revisited Childress and Backus bounds for laminar flows and considered implications for the field growth in turbulence.


{\bf Acknowledgments.}  This research was supported by the Israel Science Foundation (ISF) under grants No. 2040/17 and 366/15.

\appendix
\section{Generalised Rayleigh-Faber-Krahn inequality}
\label{appendix:A}

Here we derive the eigenmodes and eigenvalues $\lambda$ that solve Eq.~(\ref{op}). Taking divergence of the equation gives that $\nabla^2 p=0$ so that $p$ can be written as superposition of spherical harmonics that are regular at the origin, i.e. $r^lY_{lm}(\theta, \phi)$. We look for elementary solutions $\bm u_{lm}$ of Eq.~(\ref{op}) that correspond to $p=r^lY_{lm}(\theta, \phi)$. We will see that the solution allows also to find the modes with zero pressure. We consider
\begin{eqnarray}&&\!\!\!\!\!\!\!\!\!\!\!\!\!\!\!\!\!
\lambda_{lm}\bm u_{lm} \!+\!\nabla^2 \bm u_{lm}\!=\!\nabla \left(r^lY_{lm}\right)\!=\!r^{l-1}\left(l\bm Y_{lm}\!+\!\bm \Psi_{lm}\right), \label{qw}
\end{eqnarray}
where we introduced dimensionless vector spherical harmonics (VSH) \cite{sph},
\begin{eqnarray}&&\!\!\!\!\!\!\!\!\!\!\!\!\!\!\!
\bm Y_{lm}={\hat r}Y_{lm},\ \ \bm \Psi_{lm}=r\nabla Y_{lm}={\hat \theta}\partial_{\theta}Y_{lm}+\frac{{\hat \phi}\partial_{\phi}Y_{lm}}{\sin\theta},\nonumber\\&&\!\!\!\!\!\!\!\!\!\!\!\!\!\!\!
\bm \Phi_{lm}\!=\!\bm r\!\times \!\nabla Y_{lm}\!=\!-\!\nabla\!\times\! (\bm r Y_{lm})\!=\!{\hat \phi}\partial_{\theta}Y_{lm}\!-\!\frac{{\hat \theta}\partial_{\phi}Y_{lm}}{\sin\theta}, \label{vsh}
\end{eqnarray}
where $\nabla$ is the three-dimensional gradient $\nabla Y_{lm}={\hat \theta}\partial_{\theta}Y_{lm}/r+{\hat \phi}\partial_{\phi}Y_{lm}/(r\sin\theta)$. We use the multiplicative factor in the definition of spherical harmonics used in \cite{sph} where,
\begin{eqnarray}&&\!\!\!\!\!\!\!\!\!\!\!\!
\int Y_{lm} Y^*_{l'm'}d\Omega=\int_0^{\pi}\sin\theta d\theta\int_0^{2\pi}d\phi Y_{lm} Y^*_{l'm'}=\delta_{l l'}\delta_{m m'},\nonumber\\&&\!\!\!\!\!\!\!\!\!\!\!\!
Y_{lm}=\sqrt{\frac{(2l+1)}{4\pi}\frac{(l-m)!}{(l+m)!}}P_l^m(\cos\theta)\exp\left(im\phi\right).
\label{fi}
\end{eqnarray}
Thus for instance,
\begin{eqnarray}&&\!\!\!\!\!\!\!\!\!\!\!\!
Y_{10}=\sqrt{\frac{3}{4\pi}}\cos\theta,\ \ \bm \Psi_{10}=-{\hat \theta}\sqrt{\frac{3}{4\pi}}\sin\theta. \label{definition}
\end{eqnarray}
The VSH are a complete set of functions and we can expand any flow $\bm u$ as,
\begin{eqnarray}&&\!\!\!\!\!\!\!\!\!\!\!\!\!\!\!
\bm u\!=\!\sum_{l=1}^{\infty}\sum_{m=-l}^l\! \left(c^r_{lm}(r)\bm Y_{lm}\!+\!c^{(1)}_{lm}(r)\bm \Psi_{lm}\!+\!c^{(2)}_{lm}(r)\bm \Phi_{lm}\right).
\label{expansion}
\end{eqnarray}
Operating on both sides of Eq.~(\ref{expansion}) with the curl operator yields \citep{sph},
\begin{eqnarray}&&\!\!\!\!\!\!\!\!\!\!\!\!
\nabla\times \bm u=\sum_{l=1}^{\infty}\sum_{m=-l}^l \left(-\frac{l(l+1)c^{(2)}_{lm}\bm Y_{lm}}{r}
\right.\\&&\!\!\!\!\!\!\!\!\!\!\!\!\left.-\left(\frac{dc^{(2)}_{lm}}{dr}+\frac{c^{(2)}_{lm}}{r}\right)\bm \Psi_{lm}+\left(\frac{dc^1_{lm}}{dr}+\frac{c^{(1)}_{lm}}{r}-\frac{c^r_{lm}}{r}\right)
\bm \Phi_{lm}\right).\nonumber
\end{eqnarray}
Applying the curl operator once again and using the incompressibility condition that implies $\nabla^2\bm u=-\nabla\times (\nabla\times \bm u)$, we find:
\begin{eqnarray}&&\!\!\!\!\!\!\!\!\!\!\!\!
\nabla^2\bm u=\sum_{l=1}^{\infty}\sum_{m=-l}^l \left(\frac{l(l+1)\bm Y_{lm}}{r}\left(\frac{dc^{(1)}_{lm}}{dr}+\frac{c^{(1)}_{lm}}{r}-\frac{c^r_{lm}}{r}\right)
\right.\nonumber\\&&\!\!\!\!\!\!\!\!\!\!\!\!\left.+\bm \Psi_{lm}\left(\frac{d}{dr}+\frac{1}{r}\right)\left(\frac{dc^{(1)}_{lm}}{dr}+\frac{c^{(1)}_{lm}}{r}-\frac{c^r_{lm}}{r}\right)
\right.\\&&\!\!\!\!\!\!\!\!\!\!\!\!\left.
-\bm \Phi_{lm}\left[\frac{l(l+1)c^{(2)}_{lm}}{r^2}-\left(\frac{d}{dr}+
\frac{1}{r}\right)\left(\frac{dc^{(2)}_{lm}}{dr}+\frac{c^{(2)}_{lm}}{r}\right)\right]
\right).\nonumber
\end{eqnarray}
Finally, inserting the above expression for $\nabla^2\bm u$ into Eq.~(\ref{qw}) and equating coefficients of the VSH yield the following equations:
\begin{eqnarray}&&\!\!\!\!\!\!\!\!\!\!\!\!
\lambda_{lm} c^r_{lm}+\frac{l(l+1)}{r}\left(\frac{dc^{(1)}_{lm}}{dr}+\frac{c^{(1)}_{lm}}{r}-\frac{c^r_{lm}}{r}\right)
=lr^{l-1},\label{sd}\\&&\!\!\!\!\!\!\!\!\!\!\!\!
\lambda_{lm}c^{(1)}_{lm}+\left(\frac{d}{dr}+\frac{1}{r}\right)\left(\frac{dc^{(1)}_{lm}}{dr}+\frac{c^{(1)}_{lm}}{r}-\frac{c^r_{lm}}{r}\right)=r^{l-1},\nonumber\\&&\!\!\!\!\!\!\!\!\!\!\!\!
\lambda_{lm} c^{(2)}_{lm}-\frac{l(l+1)c^{(2)}_{lm}}{r^2}+\left(\frac{d}{dr}+\frac{1}{r}\right)\left(\frac{dc^{(2)}_{lm}}{dr}+\frac{c^{(2)}_{lm}}{r}\right)=0.\nonumber
\end{eqnarray}
The last equation can be written as
\begin{eqnarray}&&\!\!\!\!\!\!\!\!\!\!\!\!
\frac{d^2}{dr^2}\left(rc^{(2)}_{lm}\right)+\left(\lambda_{lm}-\frac{l(l+1)}{r^2}\right)rc^{(2)}_{lm}=0,\ \
\label{d1}
\end{eqnarray}
so that its solution that is regular at $r=0$ is given by:
\begin{equation}
c^{(2)}_{lm}=j_l(\sqrt{\lambda_{lm}} r),
\end{equation}
where $j_l(x)$ is the spherical Bessel function which is defined as the product of the Bessel function of order $l+1/2$ and $\sqrt{\pi/(2x)}$.
We omitted an arbitrary multiplicative constant. We observe that divergence of the flow is independent of $c^{(2)}_{lm}$
\begin{eqnarray}&&\!\!\!\!\!\!\!\!\!\!\!\!\!\!\!\!\!\!\!
\nabla\!\cdot\! \bm u\!=\!\sum_{l=1}^{\infty}\sum_{m=-l}^l \left(\frac{dc^r_{lm}}{dr}\!+\!\frac{2c^r_{lm}}{r}\!-\!\frac{l(l\!+\!1)c^{(1)}_{lm}}{r}\right)Y_{lm}. \label{inc}
\end{eqnarray}
Thus the incompressibility condition does not impose any restrictions on $c^{(2)}_{lm}$. We therefore find a set of eigenmodes of Eqs.~(\ref{op}) that have eigenvalues $\lambda=\lambda_{lmn}$ as determined from
\begin{eqnarray}&&\!\!\!\!\!\!\!\!\!\!\!\!\!\!\!\!\!\!\!\!\!
\bm u\!=\!\bm u^{(2)}_{lmn}\!=\!j_{l}(\sqrt{\lambda_{ln}} r)\bm \Phi_{lm},\ \ p\!=\!0,\ \ j_{l}(\sqrt{\lambda_{ln}} R)\!=\!0, \label{modes}
\end{eqnarray}
where $n$ is the radial mode number that counts zeros of the Bessel function.
Since $\bm \Phi_{00}=0$ is the trivial solution then these solutions exist for any positive integer $l$ and $-l\leq m\leq l$. The eigenvalues are independent of $m$ so that for each $l$ there is $2l+1$ degeneracy. Therefore the eigenvalues are denoted only by $l$ and $n$ while the subscript $m$ is omitted for convenience.
The smallest $\lambda_{ln}$ is attained at $l=1$ and $n=1$. We find using
\begin{eqnarray}&&\!\!\!\!\!\!\!\!\!\!\!\!\!\!\!\!\!\!\!\!\!\!
j_1(\sqrt{\lambda_{1n}} r)\!=\!
\frac{1}{ \sqrt{\lambda_{1n}} r}\left(\frac{\sin(\sqrt{\lambda_{1n}} r)}{\sqrt{\lambda_{1n}} r}\!-\!\cos(\sqrt{\lambda_{1n}} r)\right), \label{or}
\end{eqnarray}
and requiring $j_{l}(\sqrt{\lambda_{11}} R)=0$ that 
\begin{eqnarray}&&\!\!\!\!\!\!\!\!\!\!\!\!
\lambda_{11}=\frac{x^2}{R^2},\ \ \tan x=x,\ \ x\approx 4.493. \label{fx}
\end{eqnarray}
which is larger than $\pi^2/R^2$ obtained by disregarding the solenoidality constraint.
The rest of Eqs.~(\ref{sd}) read
\begin{eqnarray}&&\!\!\!\!\!\!\!\!\!\!\!\!
\frac{d(rc^{(1)}_{lm})}{dr}-c^r_{lm}=\frac{r^{l+1}}{l+1}-\frac{\lambda_{lm}r^2c^{r}_{lm}}{l(l+1)},\nonumber\\&&\!\!\!\!\!\!\!\!\!\!\!\!
\frac{d^2}{dr^2}\left(rc^{(1)}_{lm}\right)-\frac{dc^r_{lm}}{dr}=r^l-\lambda_{lm} rc^{(1)}_{lm}.
\end{eqnarray}
Taking derivative of the first equation we find that consistency of the equations demands that,
\begin{eqnarray}&&\!\!\!\!\!\!\!\!\!\!\!\!
\frac{d\left(r^2c^r_{lm}\right)}{dr}=l(l+1)rc^{(1)}_{lm}, \label{consist}
\end{eqnarray}
which agrees with the incompressibility condition, see Eq.~(\ref{inc}). Thus the solution reduces to the coupled equations,
\begin{eqnarray}&&\!\!\!\!\!\!\!\!\!\!\!\!
\frac{d(rc^{(1)}_{lm})}{dr}-\frac{r^{l+1}}{l+1}=c^r_{lm}-\frac{\lambda_{lm}r^2c^{r}_{lm}}{l(l+1)},\nonumber\\&&\!\!\!\!\!\!\!\!\!\!\!\!
\frac{d\left(r^2c^r_{lm}\right)}{dr}=l(l+1)rc^{(1)}_{lm}.\label{c1}
\end{eqnarray}
Taking derivative of the last equation and using the first,
\begin{eqnarray}&&\!\!\!\!\!\!\!\!\!\!\!\!
\frac{d^2y}{dr^2}+\left[\lambda_{lm}-\frac{l(l+1)}{r^2}\right]y=lr^{l+1}, \label{radi}
\end{eqnarray}
where $y=r^2c^r_{lm}$. We observe that $lr^{l+1}/\lambda_{lm}$ gives a partial solution of this equation. We conclude that the solution that is regular at zero is,
\begin{eqnarray}&&\!\!\!\!\!\!\!\!\!\!\!\!
c^r_{lm}=\frac{{\tilde c}^{r}_{lm} j_{l}(\sqrt{\lambda_{lm}} r)}{r}+\frac{lr^{l-1}}{\lambda_{lm}},
\end{eqnarray}
where ${\tilde c}^{r}_{lm}$ is a constant and $l>0$ (the solution with $l=0$ is not regular at the origin). We find using Eq.~(\ref{consist}),
\begin{eqnarray}&&\!\!\!\!\!\!\!\!\!\!\!\!
c^{(1)}_{lm}\!=\!\frac{\left(r^2c^r_{lm}\right)'}{l(l\!+\!1)r}
\!=\!\frac{{\tilde c}^{r}_{lm} }{l(l\!+\!1)r}\frac{d\left(rj_{l}(\sqrt{\lambda_{lm}} r)\right)}{dr}
\!+\!\frac{r^{l-1}}{\lambda_{lm}}.
\end{eqnarray}
We conclude that $\bm u_{lm}$ has the form
\begin{eqnarray}&&\!\!\!\!\!\!\!\!\!\!\!\!\!\!\!
\bm u_{lm}\!=\!\left(\frac{{\tilde c}^{r}_{l} j_{l}(\sqrt{\lambda_l} r)}{r}\!+\!\frac{lr^{l-1}}{\lambda_l}\right)\bm Y_{lm}
\nonumber\\&&\!\!\!\!\!\!\!\!\!\!\!\!\!\!\!
+\left(\frac{{\tilde c}^{r}_{l}\left(r j_{l}(\sqrt{\lambda_l} r)\right)' }{l(l+1)r}
\!+\!\frac{r^{l-1}}{\lambda_l}\right)\bm \Psi_{lm}, \label{fomrs}
\end{eqnarray}
where we omitted the subscript $m$ from ${\tilde c}^{r}_{l}$ and $\lambda_l$ as they are independent of $m$. These constants are fixed by the requirement that $\bm u_{lm}(r=R)=0$ which gives rise to an infinite series of eigenvalues $\lambda_{ln}$ where $n=1,...\infty$ is the radial mode number. The smallest eigenvalue is attained for $l=n=1$ and is found as the smallest positive solution of the coupled equations for $l=1$
\begin{eqnarray}&&\!\!\!\!\!\!\!\!\!\!\!\!
\frac{{\tilde c}^{r}_l}{R^3}\left(x\cos x-\sin x\right)=1,\\&&\!\!\!\!\!\!\!\!\!\!\!\!
\frac{{\tilde c}^{r}_l }{2 R^3}\left(\sin x\!-\!
x\cos x\!-\!x^2\sin x\right)
\!=\!1,\nonumber
\end{eqnarray}
where we introduced $\sqrt{\lambda_{11}} R=x$. The eigenvalue is found by taking the ratio of these equations. This yields that $x$ as the smallest positive root of $3\sin x-3x\cos x-x^2\sin x$ which is given by about $5.763$.

To summarise, for each $l$ and $m$ we find two sets of linearly independent eigenmodes, one with zero pressure given by Eq.~(\ref{modes}), and one with pressure equal to $r^lY_{lm}(\theta, \phi)$ given by Eq.~(\ref{fomrs}). The number of modes fits the number of components of $\bm u$ minus one due to the incompressibility constraint.  In addition since the smallest eigenvalue of the zero pressure modes, given in Eq.~(\ref{fx}) is smaller than the smallest eigenmode of the other set, we conclude that using the solenoidality condition we can tighten the bound given by Eq.~(\ref{krahn}) as given by Eq.~(\ref{kr1}) of the main text.
Our generalized Rayleigh conjecture then tells that the bound in this form holds for $V$ of any shape. It holds as equality only if $V$ is a ball and $\bm u$ is the
mode with the smallest eigenvalue that we derived above. As an example we compare this inequality with the inequality in Serrin \cite{serrin}. Serrin's inequality states
that if the domain of the flow can be enclosed in a cube with side $d$ then for any solenoidal $\bm u$ obeying $\bm u(S)=0$
\begin{eqnarray}&&\!\!\!\!\!\!\!\!\!\!\!\!\!\!
\frac{\int_V   \left(\nabla \bm u\right)^2 d\bm x}{\int_V u^2 d\bm x} \geq \frac{\alpha}{d^2},\ \ \alpha\equiv\pi^2 \frac{3+\sqrt{13}}{2}. \label{inq1}
\end{eqnarray}
For ball of radius $R$ we have $d=2R$, and Eq.~(\ref{krahn}), which is the Rayleigh-Faber-Krahn inequality, provides a tighter bound than Serrin's. In contrast, for a cube Serrin's bound is tighter. Thus solenoidality is relevant. The newly derived bound presented in Eq.~(\ref{kr1}) is significantly stronger than Serrin's (for the cube it is tighter by the factor of $1.6$).

{}

\end{document}